\documentclass[12pt]{cupconf}
\input epsf
\usepackage{graphicx}

\title{Sounding the Solar Cycle with Helioseismology: Implications for Asteroseismology}

\shorttitle{Sounding the Solar Cycle}

\author{W. J. Chaplin}

\begin{document}

\maketitle

 \section{Introduction}
 \label{sec:intro}

My brief for the IAC Winter School was to cover observational results
on helioseismology, flagging where possible implications of those
results for the asteroseismic study of solar-type stars. My desire to
make such links meant that I concentrated largely upon results for low
angular-degree (low-$l$) solar p modes, in particular results derived
from ``Sun-as-a-star'' observations (which are of course most
instructive for the transfer of experience from helioseismology to
asteroseismology). The lectures covered many aspects of
helioseismology -- modern helioseismology is a diverse field. In these
notes, rather than discuss each aspect to a moderate level of detail,
I have instead made the decision to concentrate upon one theme, that
of ``sounding'' the solar activity cycle with helioseismology. I cover
the topics from the lectures and I also include some new material,
relating both to the lecture topics and other aspects I did not have
time to cover. Implications for asteroseismology are developed and
discussed throughout.

The availability of long timeseries data on solar-type stars, courtesy
of the NASA \emph{Kepler} Mission (Gilliland et al. 2010; Chaplin et
al. 2010) and the French-led CoRoT satellite (Appourchaux et
al. 2008), is now making it possible to ``sound'' stellar cycles with
asteroseismology. The prospects for such studies have been considered
in some depth (e.g., Chaplin et al. 2007, 2008a; Metcalfe et al. 2008;
Karoff et al. 2009), and in the last year the first convincing results
on stellar-cycle variations of the p-mode frequencies of a solar-type
star (the $F$-type star HD49933) were reported by Garc\'ia et
al. (2010). This result is important for two reasons: first, the
obvious one of being the first such result, thereby demonstrating the
feasibility of such studies; and second, the period of the stellar
cycle was evidently significantly shorter than the 11-yr period of the
Sun (probably between 1 and 2\,yr). If other similar stars show
similar short-length cycles, there is the prospect of being able to
``sound'' perhaps two or more complete cycles of such stars with
\emph{Kepler} (assuming the mission is extended, as expected, to
6.5\,yr or more). The results on HD49933 may be consistent with the
paradigm that stars divide into two groups, activity-wise, with stars
in each group displaying a similar number of rotation periods per
cycle period (e.g., see B\"ohm-Vitense 2007), meaning solar-type stars
with short rotation periods -- HD49933 has a surface rotation period
of about 3\,days -- tend to have short cycle periods. We note that
Metcalfe et al. (2010) recently found another $F$-type star with a
short (1.6\,yr) cycle period (using chromospheric H \& K
data). Extension of the \emph{Kepler} Mission will of course also open
the possibility of detecting full swings in activity in stars with
cycles having periods up to approximately the length of the solar
cycle.

The rest of my notes break down as follows. Section~\ref{sec:over}
gives an introductory overview of the solar cycle, as seen in
helioseismic data. A brief history of observations of solar cycle
changes in low-angular degree (low-$l$) solar p modes is given in
Section~\ref{sec:seislow}. Then, in Section~\ref{sec:cause}, we
consider the causes of the observed changes in the mode frequencies;
and in Section~\ref{sec:params}, we discuss variations in the mode
powers and damping rates, and what the relative sizes of those changes
imply for the underlying cause.

Section~\ref{sec:sig} considers several subtle ways in which the
stellar activity cycles can affect values of, and inferences made
from, the mode parameters. Concepts are introduced using the example
of the solar cycle, and the impact it has on low-$l$ p modes observed
in Sun-as-a-star data. Implications for asteroseismic observations of
solar-type stars are then developed. We start in
Section~\ref{sec:bias} with a discussion of the impact of stellar
activity on estimates of the mode frequencies.  This is followed in
Section~\ref{sec:seprat} by a similar discussion for frequency
separation ratios. Finally, Section~\ref{sec:peak} explains how mode
peaks in the frequency-power spectrum can be ``distorted'' by stellar
cycles, rendering commonly used fitting models inappropriate.

In Section~\ref{sec:spat} we develop a simple model to illustrate the
impact on the mode frequencies of the range in latitudes covered by
near-surface magnetic activity on solar-type stars, and discuss how
the angle of inclination can affect significantly the observed
frequency shifts (because that angle affects which mode components are
visible in the observations). We also show how measurements of the
frequency shifts of modes having different angular and azimuthal
degrees may be used to make inference on the spatial distribution of
the near-surface magnetic activity on solar-type stars.

We end in Section~\ref{sec:evol} by thinking somewhat longer-term, and
consider how much low-$l$ data would be needed to measure
\emph{evolutionary} changes of the solar p-mode frequencies.

 \section{The seismic solar cycle: overview}
 \label{sec:over}

A rich, and diverse, body of observational data is now available on
temporal variations of the properties of the global solar p modes.
The signatures of these variations are correlated strongly with the
well-known 11-year cycle of surface activity. The search for temporal
variations of the p-mode properties began in the early 1980s,
following accumulation of several years of global seismic data. The
first positive result was reported by Woodard \& Noyes (1985), who
found evidence in observations made by the Active Cavity Radiometer
Irradiance Monitor (ACRIM) instrument, on board the Solar Maximum
Mission (SMM) satellite, for a systematic decrease of the frequencies
of low-$l$ p modes between 1980 and 1984.  The first year coincided
with high levels of global surface activity, while during the latter
period activity levels were much lower. The modes appeared to be
responding to the Sun's 11-year cycle of magnetic activity. Woodard \&
Noyes found that the frequencies of the most prominent modes had
decreased by roughly 1 part in 10\,000 between the activity maximum
and minimum of the cycle. By the late 1980s, an in-depth study of
frequency variations of global p modes, observed in the Big Bear data,
had demonstrated that the agent of change was confined to the outer
layers of the interior (Libbrecht \& Woodard, 1990).

Accumulation of data from the new networks and instruments has made it
possible to study the frequency variations to unprecedented levels of
detail, and has revealed signatures of subtle, structural change in
the sub-surface layers. The discovery of solar-cycle variations in
mode parameters associated with the excitation and damping (e.g.,
power, damping rate, and peak asymmetry) followed. Patterns of flow
that penetrate a substantial fraction of the convection zone have also
been uncovered as well as possibly signatures of changes in the
rotation rate of the layers that straddle the tachocline, and much
more recently evidence for a quasi-periodic 2-yr signal, superimposed
on the solar-cycle variations of the mode frequencies.


 \begin{figure*}
 \centering
 \includegraphics[width=0.8\textwidth,clip]{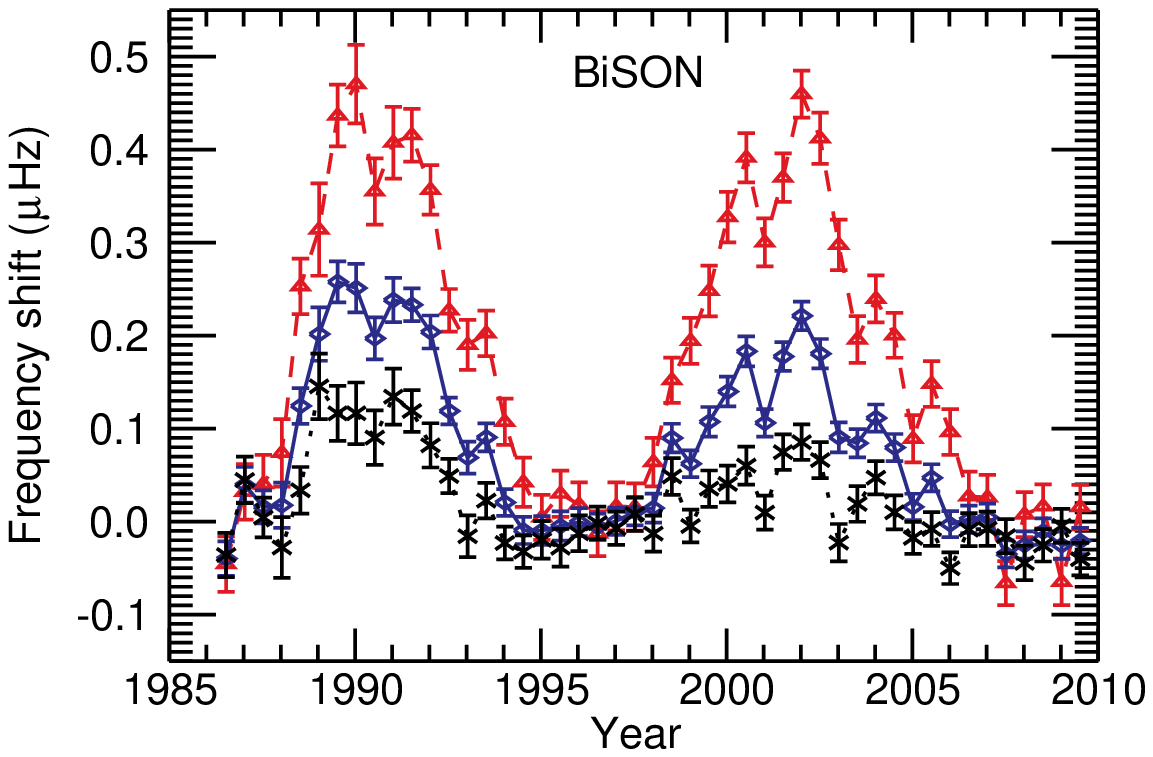}\\
 \vspace{7mm}
 \includegraphics[width=0.8\textwidth,clip]{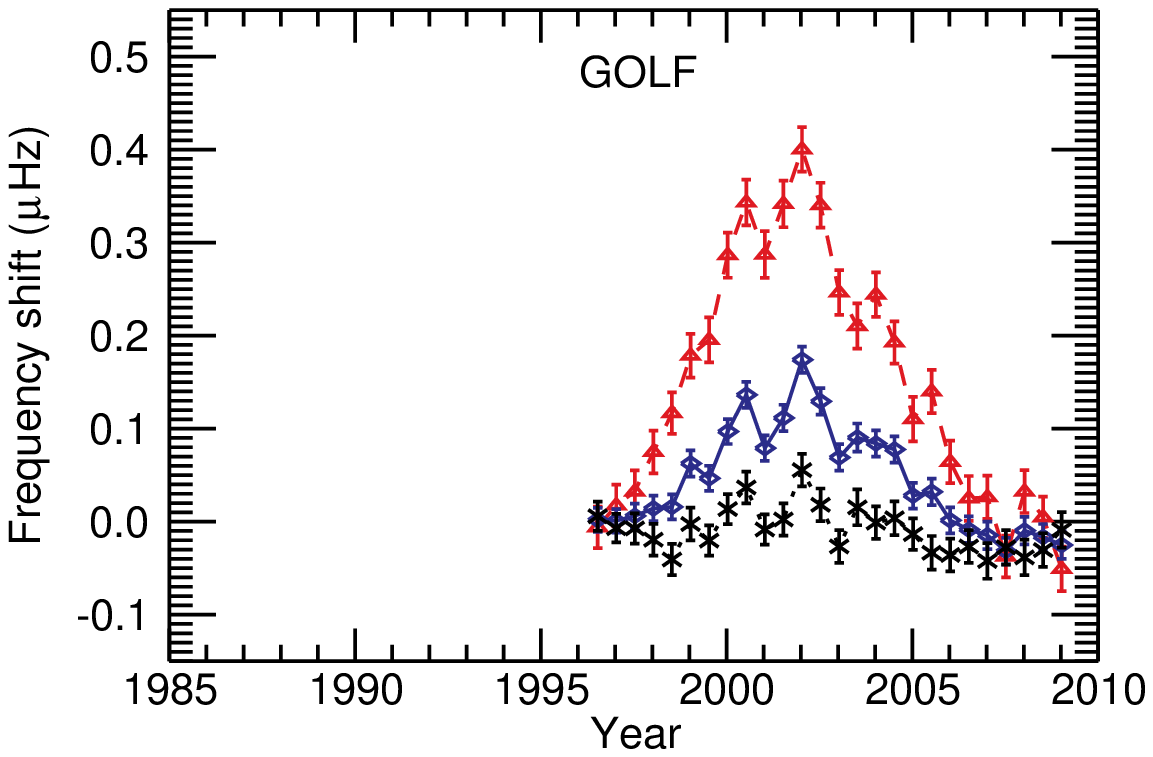}\\

 \caption{Average frequency shifts of most prominent low-$l$ solar p
 modes, as measured in BiSON data (top panel) and GOLF data (bottom
 panel) over the last two 11-yr activity cycles. The three curves show
 results for averages made over different ranges in frequency: 1880 to
 $3710\,\rm \mu Hz$ (diamonds, joined by solid line); 1880 to
 $2770\,\rm \mu Hz$ (crosses, joined by dotted line); 2820 to
 $3710\,\rm \mu Hz$ (triangles, joined by dashed line). (From Fletcher
 et al. 2010.)}

 \label{fig:bisshif}
 \end{figure*}

\begin{figure*}
 \centerline
 {\epsfxsize=12.5cm\epsfbox{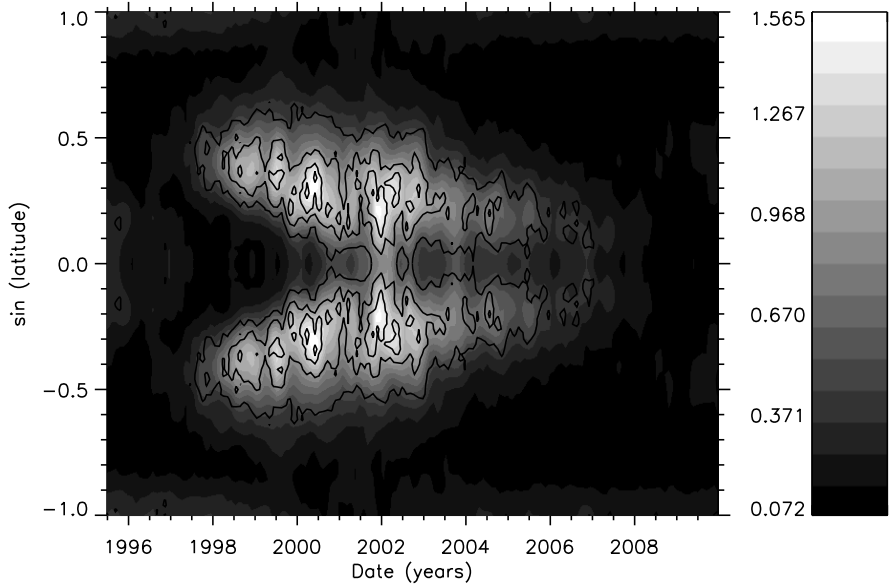}}

  \caption{Mode frequency shifts (in $\rm \mu Hz$) as a function of
   time and latitude. The values come from analysis of GONG data. The
   contour lines indicate the surface magnetic activity. (Figure
   courtesy of R.~Howe.)}

 \label{fig:numap}
\end{figure*}
\begin{figure*}
 \centerline
 {\epsfxsize=12.5cm\epsfbox{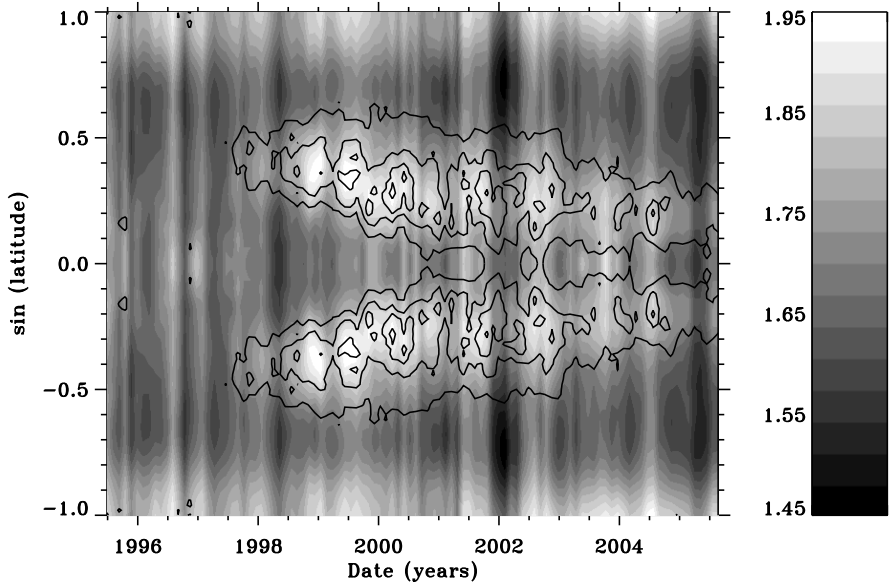}}

  \caption{Mode linewidth (in $\rm \mu Hz$) as a function of time and
   latitude. The values come from analysis of GONG data. The contour
   lines indicate the surface magnetic activity. (Courtesy of
   R.~Howe.)}

 \label{fig:widmap}
\end{figure*}

The modern seismic data give unprecedented precision on measurements
of frequency shifts. Examples of average frequency shifts for low-$l$
data are shown in Fig.~\ref{fig:bisshif}, for Sun-as-a-star data
collected by the ground-based Birmingham Solar-Oscillations Network
(BiSON), and Global Oscillations at Low Frequency (GOLF) instrument on
board the ESA/NASA Solar and Heliospheric Observatory (SOHO). From
observations of the medium-$l$ frequency shifts it is possible to
produce surface maps showing the strength of the solar-cycle shifts as
a function of latitude and time (Howe, Komm, \& Hill, 2002), like the
example shown in Figure~\ref{fig:numap} (which is made from Global
Oscillations Network Group (GONG) data).  These maps bear a striking
resemblance to the butterfly diagrams that show spatial variations in
the strength of the surface magnetic field over time. The implication
is that the frequency shift of a given mode depends on the strength of
that component of the surface magnetic field that has the same
spherical harmonic projection on the surface. Similar maps may also be
made for variations observed in the p-mode powers and damping rates
(Komm, Howe, \& Hill, 2002), which, like the frequency maps, show a
close spatial and temporal correspondence with the evolution of
active-region field (Figure~\ref{fig:widmap}).


 \subsection{The seismic solar cycle at low angular degree}
 \label{sec:seislow}

As noted above, the first evidence for activity-related changes to the
low-$l$ p modes of the Sun was reported by Woodard \& Noyes
(1985). These changes were soon confirmed (and the results extended)
by Pall\'e et al. (1989) and Elsworth et al. (1990). Almost two
decades on, there now exists an extensive literature devoted to
studies of low-$l$ mode parameter variations. Since the frequencies
are by far the most precisely determined parameters, it is hardly
surprising that analyses of the frequencies threw up the first
positive results for solar-cycle changes. Evidence for changes in mode
power followed next (Pall\'e et al. 1990a; Anguera-Gubau et al. 1992;
Elsworth et al. 1994).  The first tentative claims for changes in mode
linewidth (damping) were made by Pall\'e et al. (1990b). Subsequently,
Toutain \& Wehrli (1997) and Appourchaux (1998) provided stronger
evidence in support of an increase in damping with activity, and these
claims were confirmed beyond all doubt by Chaplin et
al. (2000). (Komm, Howe \& Hill (2000) did likewise for medium $l$
modes at about the same time.) Peak asymmetry is the most recent
addition to the list of parameters that show solar-cycle variations
(Jim\'enez-Reyes et al. 2007). Careful measurement of variations in
the powers, damping rates and peak asymmetries -- all parameters
associated with the excitation and damping -- are allowing studies to
be made of the impact of the solar cycle on the convection properties
in the near-surface layers.

Frequencies of modes below $\approx 4000\,\rm \mu Hz$ are observed to
increase with increasing activity.  For the most prominent modes at
around $3000\,\rm \mu Hz$, the size of the shift is about $0.4\, \rm
\mu Hz$ between activity minimum and maximum. Furthermore, higher
frequency modes experience a larger shift than their lower-frequency
counterparts.  This frequency dependence suggests that the
perturbations responsible for the frequency shifts are located very
close to the solar surface. The upper turning points of the modes --
which for low-degree modes are effectively independent of $l$ -- lie
deeper in the Sun for low-frequency modes than they do for
high-frequency modes: higher-frequency modes are as such more
sensitive to surface perturbations.

That the shifts do not scale like $\nu_{nl}/L$, where $\nu_{nl}$ is
mode frequency and $L=\sqrt{l(l+1)}$, rules out the possibility that
the perturbation is spread throughout a significant fraction of the
solar interior. This may be understood by thinking classically, in
terms of ray paths followed by the acoustic waves. When a wave reaches
the lower turning point of its cavity it will by definition be moving
horizontally, and its phase speed will be equal to
 \begin{equation}
 c = \frac{\omega_{nl}}{k} = \frac{2\pi \nu_{nl} R}{\sqrt{l(l+1)}}
 \propto \nu_{nl} / L,
 \label{eq:cwk}
 \end{equation}
where $k$ is the horizontal wavenumber and $R$ the outer cavity
radius. The ratio $\nu_{nl}/L$ therefore maps to the location of the
lower boundary of the cavity, and hence the cavity size. Since the
shifts do not scale like this ratio, we conclude that the perturbation
must be confined within a narrow layer, and not spread so widely that
it covers the entirety of the cavities of many of the modes.

At frequencies above $4000\,\rm \mu Hz$ the size of the shift
decreases, and also changes sign above $\simeq 4500\, \rm \mu Hz$
meaning that these very high-frequency modes suffer a reduction in
frequency as activity levels rise (Anguera-Gubau et al. 1992; Chaplin
et al. 1998; Jim\'enez-Reyes et al. 2001; Gelly et al. 2002; Salabert
et al.  2004).

Detailed comparison of the low-$l$ frequency shifts with changes in
various disc-averaged proxies of global surface activity provides
further tangible input to the solar cycle studies. This is because
different proxies show differing sensitivity to various components of
the surface activity.  While the changes in frequency are observed to
correlate fairly well with contemporaneous changes in global proxies,
the match is far from perfect. Jim\'enez-Reyes et al. (1998) were the
first to show that the relation of the frequency shifts to variations
in the proxies was markedly different on the rising and falling parts
of the 11-yr Schwabe cycle. Since the magnitudes of the shifts should
reflect the different spatial sensitivities of modes of different
angular and azimuthal degree to the time-dependent variation of the
surface distribution of the activity, a better choice for the activity
proxy would clearly be one that has been decomposed to have a similar
spatial distribution as the mode under study.  Chaplin et al. (2004a)
and Jim\'enez-Reyes et al. (2004) have shown that the sizes of the
low-$l$ shifts do indeed scale better with activity proxies that have
the same spherical harmonic projection as the modes.

Chaplin et al., (2007b) compared frequency changes in 30\,years of
BiSON data with variations in six well-known activity
proxies. Interestingly, they found that only activity proxies having
good sensitivity to the effects of weak-component magnetic flux --
which is more widely distributed in latitude than the strong flux in
the active regions -- were able to follow the frequency shifts
consistently over the three cycles.

The unusual behaviour during the most recent solar minimum (straddling
solar cycles 23 and 24) of many diagnostics and probes of solar
activity has raised considerable interest and debate in the scientific
community (e.g., see the summary by Sheeley 2010). The minimum was
unusually, and unexpectedly, extended and deep. Polar magnetic fields
were very weak, and the open flux was diminished compared to other
preceding minima.

Helioseismology has been used to probe the behaviour of sub-surface
flows during the minimum. Howe et al. (2009) found that the
equatorward progression of the lower branches of the so-called
torsional oscillations (east-west flows) was late in starting compared
to previous cycles. They flagged this delayed migration as a possible
pre-cursor of the delayed onset of cycle 24. The meridional
(north-south) flow also carries a signature of the solar cycle, which
converges toward the active-region latitudes and also intensifies in
strength as activity increases: Gonz\'alez-Hern\'andez (2010) found
that during the current minimum this component had developed to
detectable levels even before the visual onset of magnetic activity on
the solar surface.

The globally coherent acoustic properties of the recent solar minimum
have been studied extensively with low-degree p modes (by Broomhall et
al. 2009 and Salabert et al. 2009) and medium-degree p modes (Tripathy
et al. 2010). These studies have shown that while the surface proxies
of activity (e.g., the 10.7-cm radio flux) were quiescent and very
stable during the minimum, the p-mode frequencies showed much more
variability. Tripathy et al. (2010) noted further surprising behaviour
compared to the previous cycle~22-23 minimum, i.e., an apparent
anti-correlation of the p-mode frequency shifts and the surface
proxies of activity.

\begin{figure*}
 \centering
 \includegraphics[width=0.8\textwidth,clip]{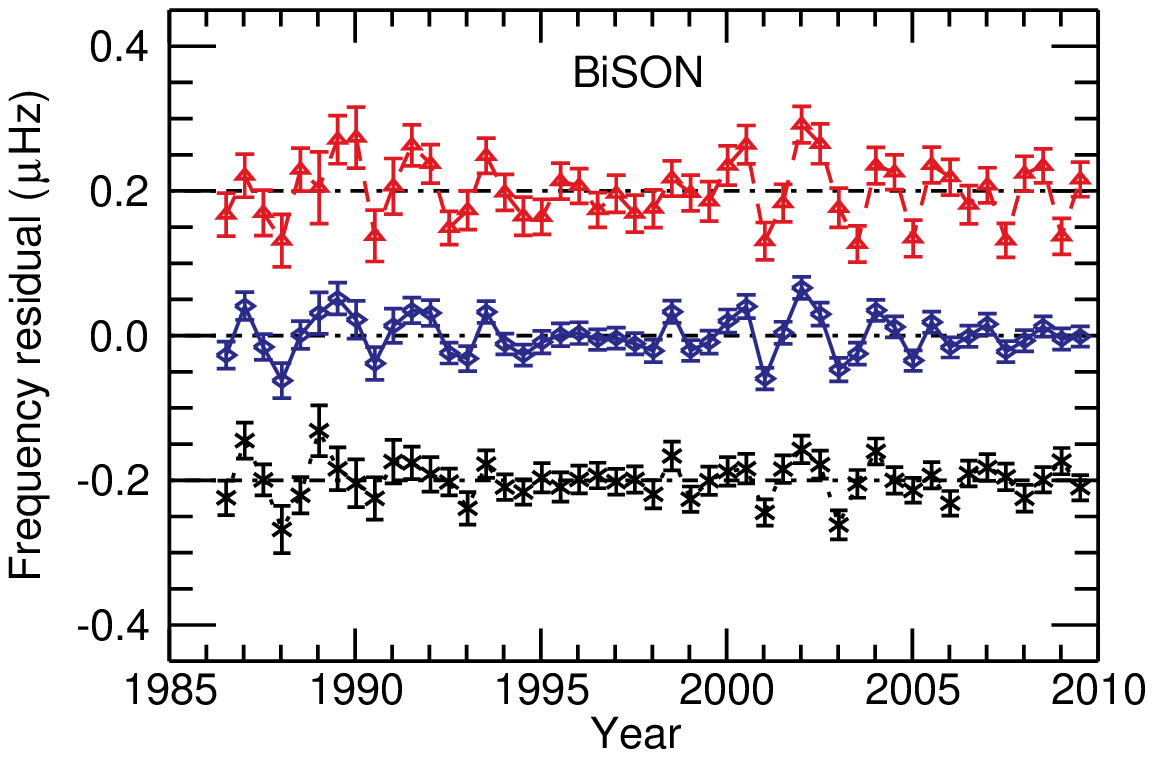}\\
 \vspace{7mm}
 \includegraphics[width=0.8\textwidth,clip]{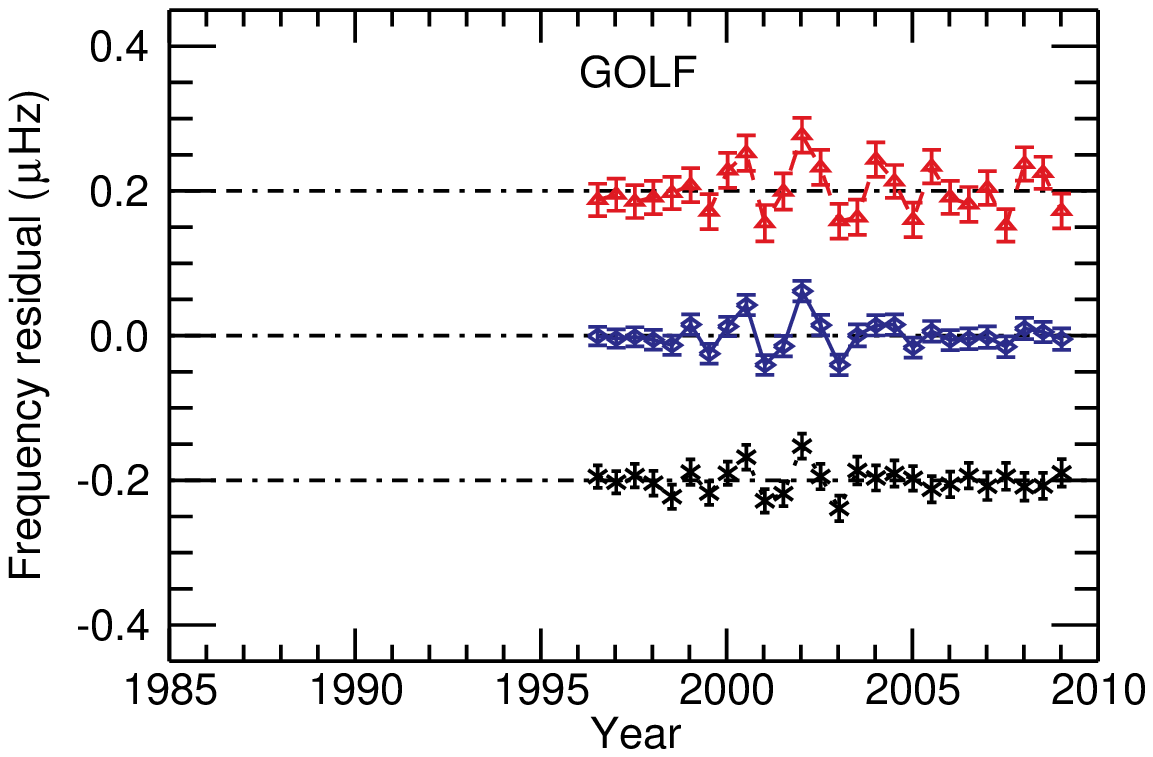}\\

 \caption{Frequency residuals that remain after the long-term
  solar-cycle variation has been removed from the average frequency
  shifts of BiSON data (top panel) and GOLF data (bottom panel).  The
  three curves in each panel show results for averages made over
  different ranges in frequency: 1880 to $3710\,\rm \mu Hz$ (diamonds,
  joined by solid line); 1880 to $2770\,\rm \mu Hz$ (crosses, joined
  by dotted line); 2820 to $3710\,\rm \mu Hz$ (triangles, joined by
  dashed line). (Plot from Fletcher et al. 2009.)}

 \label{fig:bienn1}
 \end{figure*}

Broomhall et al. (2009) had suggested the possible presence of a
quasi-biennial modulation of the frequencies of the low-degree modes,
superimposed upon the well-established $\sim 11$-yr variation of the
frequencies. This has since been confirmed by further in-depth
analysis, which reveals a signature that is consistent in the
frequencies extracted from BiSON and GOLF data (Fletcher et al. 2010),
as shown in Fig.~\ref{fig:bienn1}. This plots the frequency residuals
that remain after the long-term solar-cycle variation has been removed
from the average frequency shifts. The residuals show significant
variability, with a period of around 2\,yr, that variability being
more pronounced at times of higher surface activity.  The fact that
this biennial signature has similar amplitude in the low-frequency
\emph{and} the high-frequency modes used in this analysis suggests
that its origins lie deeper than the very superficial layers
responsible for the 11-year shifts.


Finally in this section on the frequencies, we note that from
appropriate combinations of low-$l$ frequencies Verner et al. (2006)
uncovered apparent solar-cycle variations in the amplitude of the
depression in the adiabatic index, $\Gamma_1$, in the He~{\sc ii}
zone. These variations presumably reflect the impact of the changing
activity on the equation of state of the gas in the layer, and
confirmed the findings of Basu and Mandel (2004), who used the more
numerous data available from medium-$l$ frequencies.

 \subsection{What is the cause of the frequency shifts?}
 \label{sec:cause}

Broadly speaking, the magnetic fields can affect the modes in two
ways. They can do so directly, by the action of the Lorentz force on
the gas. This provides an additional restoring force, the result being
an increase of frequency, and the appearance of new modes. Magnetic
fields can also influence matters indirectly, by affecting the
physical properties in the mode cavities and, as a result, the
propagation of the acoustic waves within them. This indirect effect
can act both ways, to either increase or decrease the frequencies.

We begin this section with a back-of-the-envelope calculation, which
makes (an admittedly) approximate prediction of the impact on the
frequency of a typical low-$l$ mode of changes in stratification
brought about by the near-surface magnetic field in sunspots. The
calculation is instructive, in that the size of the predicted
frequency shift it gives is broadly in line with the observations,
providing further evidence in support of the shifts being caused by
near-surface perturbations due to the presence of magnetic field.

As we have seen, frequencies of the most prominent modes increase with
increasing magnetic activity. Here, we consider the indirect effect of
the near-surface magnetic field on the near-surface properties, and
hence the frequencies of these modes. If magnetic fields modify the
surface properties in such a way as to reduce the effective size of
the mode cavity, the required increase of frequency will result.

In regions of strong magnetic field -- such as those occupied by
sunspots -- there will be a gas pressure deficit (assuming those
regions to be in pressure equilibrium with their field-free
surroundings). This is because gas and magnetic pressure combine
within the magnetic regions, while only gas pressure acts in the
field-free regions. Sunspots are also characterized by a reduced
temperature, relative to the surroundings. The central part of the
spot therefore exhibits a lower pressure, temperature and density than
the surroundings, resulting in the so-called Wilson Depression. Values
of pressure, temperature and density found at the surface in the
field-free plasma are only reached at some depth beneath the surface
in strong-field regions. Recent measurements suggest that for a
sunspot the typical size of this depression is about 1000\,km (Watson
et al. 2009).

Let us assume that 1000\,km corresponds approximately to the amount,
$\delta R$, by which the mode cavities are reduced in size beneath
sunspots. The fractional area of the solar surface occupied by
sunspots reaches $\sim 0.5\,\%$ at modern cycle maxima. We therefore
obtain a net, surface-averaged estimate of $\delta R$ via
 \[
 \left< \delta R \right> \approx 0.5\,\% \times 1000\,{\rm km} \approx
 5\,\rm km.
 \] 
The sound speed, $c$, at the solar surface is approximately $10\,\rm
km\,s^{-1}$, implying a reduction in the travel time in the mode
cavity, due to the ``shrinkage'' at the surface, of
 \[
 \delta T \approx 5\,\rm km / 10\,km\,s^{-1} \approx
 0.5\,s.
 \] 
The travel time across the cavity, $T$, is related to the large
frequency separation, $\Delta\nu$, via:
 \begin{equation}
 \Delta\nu \simeq \left( 2 \int_0^R \frac{dr}{c} \right)^{-1} \simeq (2T)^{-1}.
 \label{eq:cav}
 \end{equation}
For the Sun, $\Delta\nu = 135\,\rm \mu Hz$, implying $T \approx
3700\,\rm s$. The fractional change (reduction) in $T$ is therefore:
 \[
 \delta T / T \simeq 0.5/3700 \simeq 1.4 \times
 10^{-4}.
 \]
Since $\delta\nu/\nu = -\delta T/T$, the predicted increase in
frequency of a mode at $\approx 3000\,\rm \mu Hz$ will be
 \[
 \delta\nu = 3000 \times 1.4 \times 10^{-4} \simeq 0.4\,\rm \mu Hz.
 \]
This estimate is of a very similar size to the observed frequency
shifts.

What does detailed modelling suggest? Perhaps the most significant
contribution of recent years in this area is that of Dziembowski and
Goode (2005). Their results suggest that the indirect effects dominate
the perturbations, and that the magnetic fields are too weak in the
near-surface layers for the direct effect to contribute significantly
to the observed frequency shifts.  However, Dziembowski and Goode also
found some evidence to suggest that the direct effect may play a more
important r\^ole for low-frequency modes, at depths beneath the
surface where the magnetic field is strong enough to give a
significant direct contribution to the frequency shifts.  We now go on
to explain how dependence of the frequency shifts on mode inertia and
mode frequency can tell us something about the location and nature of
the perturbations.

\begin{figure*}
 \centerline
 {\epsfxsize=11.0cm\epsfbox{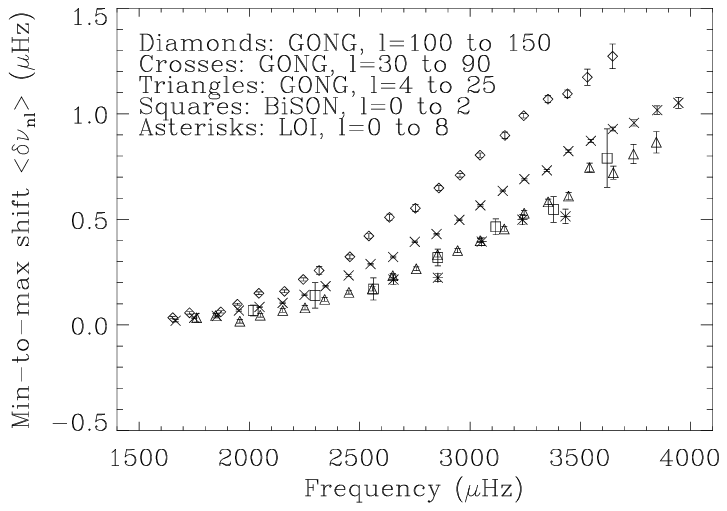}}
 \centerline
 {\epsfxsize=11.0cm\epsfbox{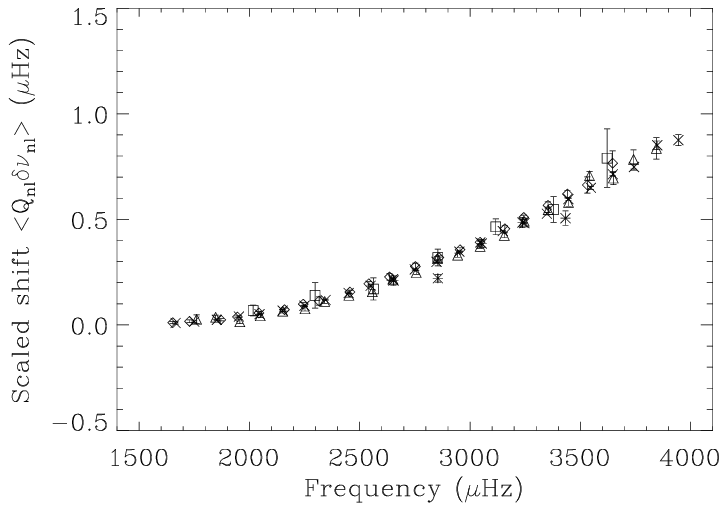}}

  \caption{Top panel: Frequency shifts, $\delta\nu$, plotted as a
  function of mode frequency, for results from different instruments
  over different ranges in $l$ (see plot annotation). Bottom panel:
  Scaled frequency shifts, after multiplication by the inertia ratio
  $Q_{nl}$. (From Chaplin et al. 2001.)}

 \label{fig:ldep}
\end{figure*}

We begin by noting that when the frequency shifts are multiplied by
the mode inertia, and then normalized by the inertia of a radial mode
of the same frequency, the modified shifts are found to be a function
of frequency alone. This in effect removes any $l$ dependence of the
shifts (at fixed frequency, the higher the $l$, the larger is the
observed frequency shift). The observed $l$ dependence may be
understood in terms of, for example, a physical interpretation of the
mode inertia. The normalized inertia, $I_{nl}$, may be defined
according to (Christensen-Dalsgaard \& Berthomieu 1991):
 \begin{equation}
 I_{nl} = {\rm M}_{\odot}^{-1} \int_{V} |\xi|^2 \rho dV = 
          4\pi {\rm M}_{\odot}^{-1} \int_0^{R} |\xi|^2 \rho r^2 dr = 
          M_{nl} / \rm M_{\odot}
 \label{eq:modem}
 \end{equation}
where $\xi$ is the (surface-normalized) displacement associated with
the mode, and the integration is performed over the volume $V$ of the
Sun, which has mass $\rm M_{\odot}$. The mode mass $M_{nl}$ is
therefore the interior mass affected by the perturbations associated
with the mode. As $l$ increases, so $M_{nl}$ decreases, and the more
sensitive a mode will be to a near-surface perturbation of a given
size (giving a larger frequency shift). One may therefore render the
shifts $l$ independent by multiplying them by the inertia ratio
$Q_{nl}$ (Christensen-Dalsgaard \& Berthomieu 1991), which is given
by:
 \begin{equation}
 Q_{nl} = I_{nl} / \bar{I}(\nu_{nl})
 \label{eq:q}
 \end{equation}
Multiplication of the raw shifts by $Q_{nl}$ is indeed seen to collapse
the shifts of different $l$ onto a single curve. As shown in
Fig.~\ref{fig:ldep}, this then allows one to combine data spanning a
range in $l$, which reduces errors, giving tighter constraints on
frequency dependence of the shifts.


Chaplin et al. (2001) studied in detail the frequency dependence of
the inertia-ratio-corrected shifts, $\delta\nu_{nl} Q_{nl}$, of both
low-$l$ modes and medium-$l$ modes up to $l=150$. They fitted these
data to a power law of the form
 \begin{equation}
 \delta\nu_{nl} Q_{nl} = \frac{c}{I_{nl}} \nu_{nl}^\alpha,
 \label{eq:dnunu}
 \end{equation}
where the power-law index is $\alpha$ and $c$ is a constant. They
found that $\alpha \simeq 2$ for $\nu \ge 2500\,\rm \mu Hz$, while
$\alpha$ is approximately zero for $\nu < 2500\,\rm \mu
Hz$. Rabello-Soares et al. (2008) repeated the exercise for medium-$l$
and high-$l$ modes. They extracted very similar behaviour, and,
because they had more medium- and high-$l$ data at low frequencies,
they were able to extend the analysis to $\nu < 2000\,\rm \mu Hz$
where they found that $\alpha$ appears to change sign and go
negative. The high-$l$ modes provide important information since they
are confined in the layers close to the surface where the physical
changes responsible for the frequency shifts are also located.

What do the values of $\alpha$ imply? If the perturbation is located
within the photosphere (but is confined to extend over an extent no
longer than one pressure scale height), then we might expect $\alpha
\simeq 3$ (e.g., Libbrecht \& Woodard 1990). If instead the
perturbation extends beneath the surface, the frequency dependence will
be weaker and $\alpha$ will get smaller (Gough 1990). The observed
values of $\alpha$ in the above mentioned frequency regimes are both
less than 3, suggesting that the perturbations cannot arise solely in
the photosphere. Since $\alpha$ is smaller for the lower-frequency
modes, this suggests that the perturbation extends to greater depths
the lower in frequency one goes. This is consistent with the
inferences made by Goode \& Dziembowski (2005).

 \subsection{Solar-cycle variations of mode power and mode damping}
 \label{sec:params}

Variations in mode damping have now been uncovered in all the major
low-$l$ datasets (see Chaplin 2004 and references therein).  The trend
found is an increase of about 20\,per cent between activity minimum
and maximum, with some suggestion of the variations being peaked in
size at about $\simeq 3000\,\rm \mu Hz$. Mode powers are at the same
time observed to decrease by about the same fractional amount, while
the mode heights decrease by twice the amount. Use of the analogy of a
damped, randomly forced oscillator is particularly instructive for
understanding these results.

We define our oscillator according to the equation:
 \begin{equation}
 \ddot{x}(t) + 2\eta \dot{x}(t) + (2\pi\nu_0)^2 x(t) = K\delta(t-t_0),
 \label{eq:osc}
 \end{equation}
where $x(t)$ is the displacement, $\nu_0$ the natural frequency of the
oscillator, $\eta$ the linear damping constant, and $K$ is the
amplitude of the forcing function (assumed to be a random Gaussian
variable), with ``kicks'' applied at times $t_0$, $\delta(t-t_0)$
being the delta function.

Provided that ${\cal F}(\nu)$ -- the frequency spectrum of the forcing
function -- is a slowly-varying function of $\nu$, and $\eta <<
2\pi\nu_{0}$, the power spectral density (PSD) in the frequency domain
will be a Lorentzian profile, i.e.,
 \begin{equation}
 {\rm PSD}(\nu) \propto {\cal F}(\nu) \left( 1 + \left(
 \frac{\nu-\nu_0}{\eta/2\pi} \right)^2 \right)^{-1}.
 \label{eq:oscpow}
 \end{equation}
This holds for both the spectrum of the displacement, $x(t)$, and the
velocity, $\dot{x}(t)$. The FWHM of the peak in cyclic frequency is
 \begin{equation}
 \Delta = \frac{\eta}{\pi}.
 \label{eq:fwhm}
 \end{equation}
The maximum power spectral density, or height, of the resonant peak is:
 \begin{equation}
 H \propto \frac{{\cal F}(\nu)}{\eta^2}.
 \label{eq:H}
 \end{equation}
The total mean-square power (variance in the time domain) is
proportional to peak height times width, i.e., $P \propto H \Delta$,
so that
 \begin{equation}
 P \propto \frac{{\cal F}(\nu)}{\eta}.
 \label{eq:P}
 \end{equation}
The energy (kinetic plus potential) of a resonant mode with associated
inertia $I$ is given by:
 \begin{equation}
 E = M P.
 \label{eq:E}
 \end{equation}
The rate at which energy is supplied to (and dissipated by) the modes,
$dE/dt$, is readily derived by again having recourse to the oscillator
analogy.  The amplitude of the oscillator is attenuated in time by the
factor $\exp{(-\eta t)}$, and its energy is proportional to the
amplitude squared. Hence, we may write
 \[ 
  E = ({\rm constant}) \times \exp{(-2\eta t)}.
 \]
It
follows that:
 \[ 
 \log{E} = -2\eta t + \log{\rm (constant)},
 \]
and taking derivatives:
 \begin{equation} 
 dE/dt = -2\eta E.
 \label{eq:Edot1}
 \end{equation}
If we combine Equations~\ref{eq:P},~\ref{eq:E} and~\ref{eq:Edot1}, we
have:
 \begin{equation} 
 dE/dt = \dot{E} \propto - {\cal F}(\nu) I.
 \label{eq:Edot2} 
 \end{equation}

We may attempt to measure solar-cycle variations in $\Delta$, $H$,
$P$, $E$ and $\dot{E}$.  The parameters that are usually extracted
directly by the observers are the widths ($\Delta$) and heights ($H$)
of the peaks in the frequency power spectrum. The mode powers, $P$,
are then estimated from
 \begin{equation}
 P = \frac{\pi}{2} \Delta H
 \label{eq:height}
 \end{equation}
(i.e., area under Lorentzian). The $E$ may be computed from the $P$
 using model computed inertias, $I$. We note that fractional changes
 in $E$ will be the same as those in $P$ (assuming the $I$ do not show
 any significant changes).  Finally, $\dot{E}$ follows from
 \begin{equation}
 \dot{E} = -\pi E \Delta = -\pi I P \Delta.
 \label{eq:dote}
 \end{equation}

If solar-cycle variations are uncovered in the above parameters, what
do they imply for the underlying cause (or causes)? If we think in
terms of the basic oscillator parameters -- the damping rate $\eta$
and the forcing function ${\cal F}(\nu)$ --
Equations~\ref{eq:fwhm},~\ref{eq:H},~\ref{eq:P},~\ref{eq:E}
and~\ref{eq:Edot2} then imply the following:
 \begin{equation}
 \delta \Delta\nu / \Delta\nu = \delta \eta / \eta,
 \end{equation}
 \begin{equation}
 \delta H / H = \delta {\cal F}(\nu)/ {\cal F}(\nu) -2\delta \eta / \eta,
 \end{equation}
 \begin{equation}
 \delta P / P = \delta E / E = \delta {\cal F}(\nu)/ {\cal F}(\nu) -\delta \eta / \eta,
 \end{equation}
and
 \begin{equation}
 \delta \dot{E} / \dot{E} = \delta {\cal F}(\nu)/ {\cal F}(\nu).
 \end{equation}

Observations of p modes at low $l$ are consistent with $\delta \dot{E}
/ \dot{E} = 0$, which in turn implies that $\delta {\cal F}(\nu)/
{\cal F}(\nu) = 0$: the acoustic forcing of the low-$l$ modes remains,
on average, constant over the cycle. Any changes that are observed in
the other parameters must therefore arise from $\delta \eta / \eta$
alone, and we have that:
 \begin{equation}
 \delta \Delta\nu / \Delta\nu = - \delta P / P = - \delta H / 2H,
 \label{eq:oscobs1}
 \end{equation}
The observed changes in $\Delta$, $P$ and $H$ follow these ratios. We
may therefore infer that changes to these parameters are in all
likelihood the result of changes to the damping of the low-$l$ modes.
Houdek et al. (2001) noted that the high Rayleigh number of the
magnetic field could modify the preferred horizontal length-scale of
the convection, which in turn could modify the damping rates. Their
numerical calculations of the implied changes for low-$l$ modes show
reasonable agreement with the observations.

 \section{Subtle signatures of stellar activity and stellar-cycle variations}
 \label{sec:sig}

Asteroseismic observations of stars will, for the foreseeable future,
be limited to obtaining data on low-$l$ modes because of the
cancellation effects resulting from unresolved observations of stellar
discs. Moreover, the angle of inclination, $i$, offered by a star also
determines which of the low-$l$ components will have non-negligible
visibility in the data. The visibility of the radial ($l=0$) modes is
not affected by $i$, but for the non-radial modes the visibility,
${\cal E}(i)_{lm}$ (in power) of a given ($l,m$) component is assumed
to go like (e.g., Gizon \& Solanki 2003):
 \begin{equation}
 {\cal E}(i)_{lm} \propto \frac{(l-|m|)!}{(l+|m|)!} 
 \left(P_l^{|m|}(\cos i\right)^2,
 \label{eq:vis}
 \end{equation}
where $P_l^{|m|}$ are Legendre polynomials. The fact that some
components may in effect be missing from the data has important
implications for the analysis. There are some obvious implications,
e.g., it will not be possible to measure frequency splittings of
non-radial modes if $i$ is close to 90\,degrees -- so that the star is
viewed rotation-pole on -- since only the zonal components ($m=0$)
will have non-negligible visibility. In this section we highlight a
more subtle aspect, which relates to the fact that estimates of the
frequencies of non-radial modes are affected by the interplay between
the effects of near-surface magnetic activity and the relative
visibility (or absence) of the mode components. This has consequences
for inferences made on solar-type stars. We first introduce and
explain the problems using ``Sun-as-a-star'' helioseismic observations
as a test case, before considering the wider implications for
asteroseismology. We return later, in Section~\ref{sec:spat} to
discuss the implications for estimation of average frequency shifts of
solar-type stars.

 \subsection{Explanation and description of frequency bias}
 \label{sec:bias}

Extant Sun-as-a-star observations are made from a perspective in which
the plane of the rotation axis of the Sun is nearly perpendicular to
the line-of-sight direction (i.e., $i$ is always close to
90\,degrees).  This means that only components with $l+m$ even have
non-negligible visibility. As we shall explain below, the impact of
surface magnetic activity on the arrangement of the components in
frequency means it may not then be possible to estimate the true
frequency centroid of the multiplet. The frequency centroids carry
information on the spherically symmetric component of the internal
structure, and are the input data that are required, for example, for
hydrostatic structure inversions.

In the complete absence of the near-surface activity, the $l+m$ odd
components which are ``missing'' from the Sun-a-as-a-star data would
be an irrelevance. All mode components would be arranged symmetrically
in frequency, meaning centroids could be estimated accurately from the
subset of visible components. A near-symmetric arrangement is found at
the epochs of modern solar-cycle minima (Chaplin et
al. 2004b). However, when the observations span a period having medium
to high levels of activity -- as a long dataset by necessity must --
the arrangement of components will no longer be symmetric. The
frequencies given by fitting the Sun-as-a-star data will then differ
from the true centroids by an amount that is sensitive to $l$. The $l$
dependence arises because in the Sun-as-a-star data modes of different
$l$ comprise visible components having different combinations of $l$
and $m$; and these different combinations show different responses (in
amplitude and phase) to the spatially non-homogeneous surface activity
(i.e., as observed in the active bands of latitude for solar-type
stars).


 \begin{figure*}
\centerline
 {\epsfxsize=14.0cm\epsfbox{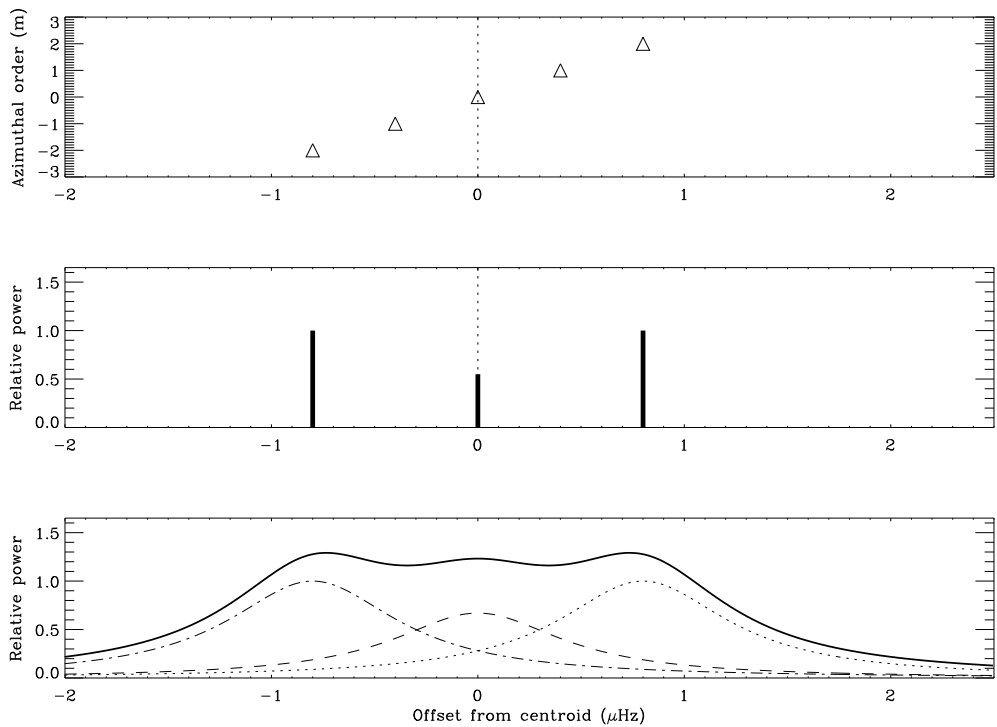}}

 \caption{Top panel: Placement in frequency, relative to the frequency
 centroid of all $2l+1$ components (abscissa), of the various $m$
 (labelled on ordinate) of an $l=2$ mode during an epoch of negligible
 solar surface activity. The solid vertical line marks the location of
 the centroid. Middle panel: placement in frequency (abscissa) and
 relative visibility in the frequency power spectrum (height on
 ordinate) of the three mode components that would have non-negligible
 visibility in the Sun-as-a-star data.  Bottom panel: Lorentzian peak
 profiles expected of an $l=2$ mode at the centre of the p-mode
 spectrum (where peak linewidths are around $1\,\rm \mu Hz$).}

 \label{fig:bias1a}
 \end{figure*}


 \begin{figure*}
\centerline
 {\epsfxsize=14.0cm\epsfbox{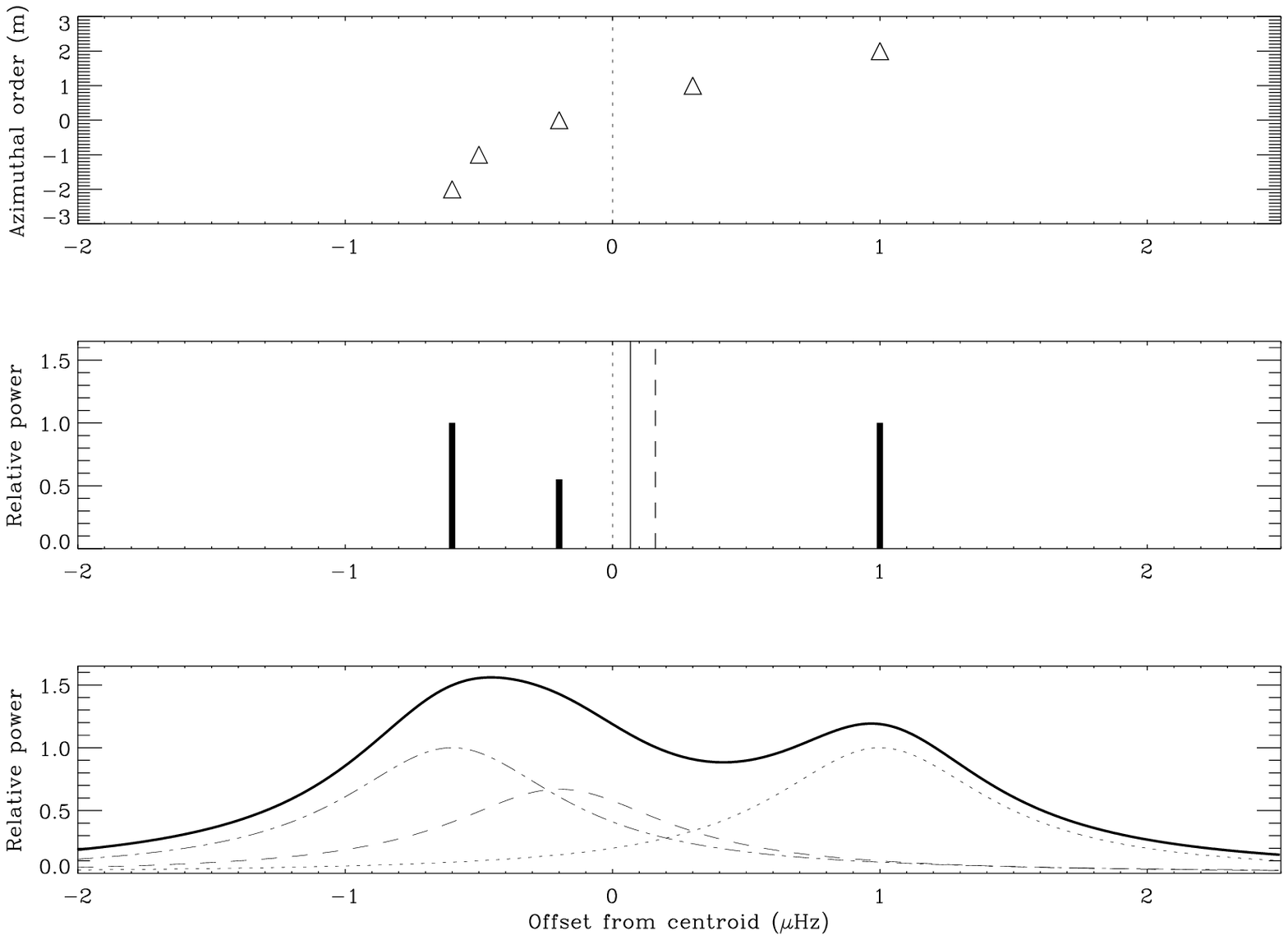}}

 \caption{Same as Fig.~\ref{fig:bias1a}, but during an epoch with
 activity levels corresponding to the maximum of a typical (modern)
 solar cycle. The solid and dashed lines in the middle panel mark
 estimates of the $l=2$ frequency that might be derived from the
 Sun-as-a-star data (see text).}

 \label{fig:bias1b}
 \end{figure*}


These effects are illustrated in Figs.~\ref{fig:bias1a}
and~\ref{fig:bias1b}. Fig.~\ref{fig:bias1a} is a schematic plot of the
arrangement in frequency of the components of an $l=2$ mode, under
circumstances where activity levels are assumed to be negligible. The
top panel shows the arrangement in frequency of the various $m$ (as
labelled on the ordinate), assuming a rotational frequency splitting
of $0.4\,\rm \mu Hz$ between adjacent $m$. The dotted line marks the
frequency centroid. The middle panel shows schematically both the
placement in frequency (abscissa) and relative visibility in the
frequency power spectrum (height on ordinate) of the three mode
components that would have non-negligible visibility in the
Sun-as-a-star data.  The bottom panel shows the Lorentzian peak
profiles expected of a mode at the centre of the p-mode spectrum
(where peak linewidths are around $1\,\rm \mu Hz$).

Fig.~\ref{fig:bias1b} shows the situation at an epoch coinciding with
high levels of solar activity. This activity is most prominent in the
active latitudes. The $|m|=2$ components are more sensitive to
acoustic perturbations due to activity that arise in these latitudes
than is the $m=0$ component (as a result of the spatial distribution
of the relevant spherical harmonic components). The result is a
distortion of the symmetric arrangement in frequency of the multiplet,
because the different $|m|$ experience a frequency shift of different
size.

The non-symmetric arrangement of the components is very evident in
Fig.~\ref{fig:bias1b}. Two lines have been plotted in the middle
panel, which were not shown in Fig.~\ref{fig:bias1a}. The solid line
marks the (unweighted) average frequency of the components showing
non-negligible visibility. It is clearly offset, or biased, by about
$0.07\,\rm \mu Hz$, from the centroid (dotted line).  The dashed line
marks the actual, expected location of the ``Sun-as-a-star'' frequency
(offset from the centroid by more than $0.1\,\rm \mu Hz$). It differs
from the unweighted average of the three visible components because
the ``peak bagging'' procedures used to estimate the mode frequencies
are influenced by the relative heights of the different $m$. Here, the
outer $|m|=2$ components carry greater weight in determining the
fitted $l=2$ frequency than does the weaker $m=0$ component, thereby
shifting the estimated frequency from the unweighted average.

It can be quite hard to estimate, to any reasonable accuracy and
precision, the locations in frequency of each component individually
(which, as noted above, would help to reduce the offset between the
estimated frequency and the true centroid because it would be possible
to compute the unweighted average frequency).  Frequency estimation
must contend with the fact that within the non-radial multiplets the
various $m$ lie in such close frequency proximity to one another that
suitable models which seek to describe the characteristics of the $m$
present must usually be fitted to the components simultaneously.

Is there any way to get around the problem, and in some way
``correct'' the Sun-as-a-star frequencies for the frequency bias?  Two
methods have been applied successfully. The first relies on the
availability of contemporaneous resolved-Sun helioseismic data --
which are sensitive to all $m$ components -- to in effect map the
strength and spatial distribution of the acoustic asphericity arising
from the surface activity over the epoch in question. Chaplin et
al. (2004c) and Appourchaux \& Chaplin (2007) show how to make the
correction, using the so-called ``even $a$ coefficients'' from fits to
the resolved-Sun frequencies (the $a$ coefficients are discussed later
in this section). This approach is of course currently not viable for
other solar-type stars. Fortunately, the second method in principle
is: here, the same dataset that is used to produce the frequency
estimates is also used to help calibrate the corrections that are
needed to ``clean'' the frequencies. Implicit in the procedure is the
assumption that at epochs coinciding with cycle minima, any frequency
asymmetry in the mode multiplets is negligible. This assumption turns
out to be reasonable for the Sun, as noted previously (Chaplin et
al. 2004b).

The second method is described at length in Broomhall et
al. (2009). In sum, the long dataset is divided into smaller subsets,
and the frequencies of each subset are estimated. Tried and trusted
procedures are then used to measure the frequency shifts of different
modes from one subset to another. The frequency shift of any given
mode may be measured relative to its average frequency over the long
dataset, or, for example, its frequency in the subsets (or subset)
which coincide(s) with the cycle minimum (the choice has subtle
consequences for propagation of the frequency uncertainties, e.g., see
Chaplin et al. 2007b). It is common practice, at least for the low-$l$
data, to average shifts over several modes to reduce uncertainties on
the shifts. By making an average over many orders in $n$, one can
track the average frequency shift from subset to subset. By making an
average over a few (say two, or three) orders in $n$ it is possible to
constrain the dependence of the shifts on frequency (or $n$).

In the next stage of the procedure the frequency shifts are
parametrized. Sun-as-a-star procedures have parametrized the shifts
as a linear function of one of the global proxies of magnetic
activity, the 10.7-cm radio flux (Tapping \& De~Tracey 1990) being a
favoured choice. The mode-averaged frequency shifts are fitted to a
linear function of the 10.7-cm flux (averaged over the same periods as
the seismic data). This linear model then suffices to predict the
expected average frequency shift for any epoch, given the 10.7-cm flux
for that period. Expected shifts for individual modes are then
computed using a best-fitting polynomial, which describes the measured
frequency shifts as a function of frequency, relative to the average
frequency shift. The predicted frequency shifts may then be subtracted
from the raw frequencies (the latter measured from the full
timeseries) to yield a set of corrected frequencies, commensurate with
negligible surface activity.  The procedure yields frequencies that
are not subject to the bias discussed above, since that bias is
assumed, \emph{by definition}, to be missing at low activity.

Application of the procedure, verbatim, to other stars would of course
demand that contemporaneous activity data be available (e.g.,
chromospheric H \& K data). Such data may be hard to obtain in
sufficient quantities on a large number of stars. One might instead
use measures of variability in the timeseries -- arising from
rotational modulation of surface activity -- to form a suitable proxy
(Basri et al. 2010).  It may instead be possible to parameterize the
frequency shifts by fitting non-linear functions in time (i.e., to
describe the observed shifts as functions of time), hence
circumventing the need to have complementary activity data.

How large might the frequency bias be for other stars? The magnitude
of the bias is governed to first order by the size of the acoustic
asphericity, arising from the magnetic activity, and the angle of
inclination of the star. Second-order effects come in through the
interplay between the mode properties and the peak-bagging procedures
used to estimate the frequencies (e.g., see discussion in Ballot et
al. 2008 on correlations between different parameters).

\begin{figure*}
\centerline
 {\epsfxsize=12.5cm\epsfbox{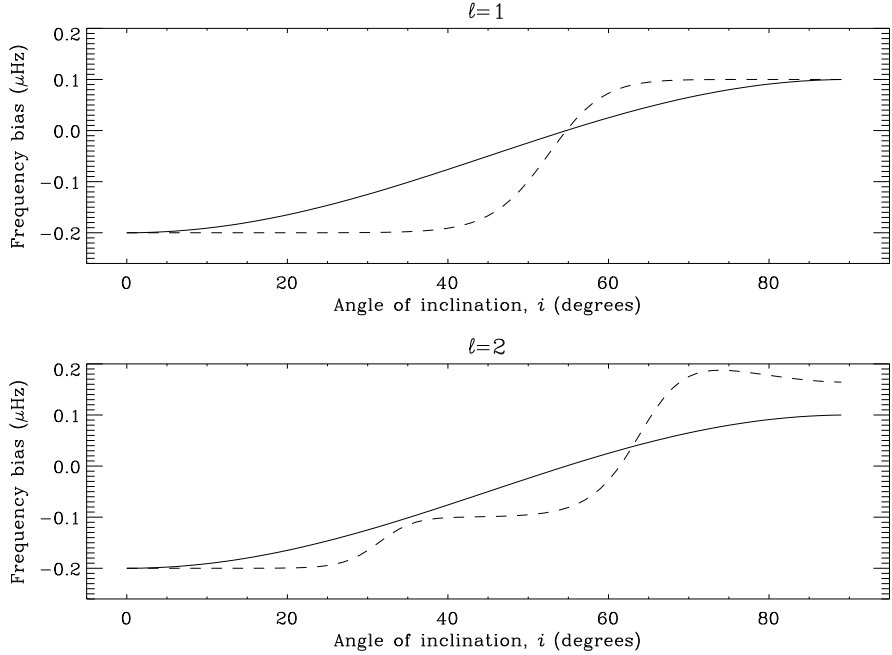}}

 \caption{Expected frequency bias for Sun-as-a-star observations made
 for different angles of inclination, $i$ (see text for explanations
 of solid and dashed lines).}

 \label{fig:bias2}
 \end{figure*}

Fig.~\ref{fig:bias2} shows the predicted bias from the first-order
effects.  Here, we assumed an acoustic asphericity similar to that
observed on the Sun, as experienced by $l=1$ and $l=2$ modes at the
centre of the solar p-mode spectrum. The predictions therefore in
principle show the bias that would be expected in the most prominent
low-$l$ modes if the Sun were to be observed at different angles of
inclination, $i$. Two predictions are plotted in each panel. We first
explain how we derived the solid-line prediction (using the $l=2$ case
as an example).

The full set of $m$ frequencies of a mode may be described by a
polynomial expansion of the form
 \begin{equation}
 \nu_{nlm} = \nu_{\rm cen} + \sum_{j=1}^{j=2l} a_j(n,l) l {\cal P}^j_l(m),
 \label{eq:acoeefs}
 \end{equation}
where $\nu_{\rm cen}$ is the frequency centroid of the mode, and
${\cal P}^j_l(m)$ are polynomials related to the Clebsch-Gordan
coefficients (Ritzwoller \& Laveley 1991). An expansion of this type
is commonly used to fit the frequencies observed in resolved-Sun data,
where all $m$ components are available. The even $a_j$ describe
perturbations that are non-spherically symmetric in nature, e.g., the
acoustic asphericity from activity, while the odd $a_j$ coefficients
describe spherically symmetric contributions to the frequency
splittings (rotation). Here, for simplicity, we assume that for the
low-$l$ modes the $a_2$ term dominates other terms in the description
of the asphericity, and that the $a_1$ term dominates other terms in
the description of the rotation. The required ${\cal P}^j_l(m)$ are:
 \begin{equation}
 {\cal P}^1_l(m) = m/l
 \label{eq:p1}
 \end{equation}
and
 \begin{equation}
 {\cal P}^2_l(m) = \frac{6m^2-2l(l+1)}{6l^2-2l(l+1)}.
 \label{eq:p2}
 \end{equation}\\
We may then write the frequencies of each of the components explicitly
as:
 \begin{equation}
 \nu_{n20} = \nu_{\rm cen} - 2a_2,
 \label{eq:nu20}
 \end{equation}
and
 \[
 \nu_{n21} = \nu_{\rm cen} + a_1 - a_2,
 \]
 \begin{equation}
 \nu_{n2\,-1} = \nu_{\rm cen} - a_1 - a_2,
 \label{eq:nu21}
 \end{equation}
and
 \[
 \nu_{n22} = \nu_{\rm cen} + 2a_1 +  2a_2,
 \]
 \begin{equation}
 \nu_{n2\,-2} = \nu_{\rm cen} - 2a_1 +  2a_2.
 \label{eq:nu22}
 \end{equation}

The solid line prediction in Fig.~\ref{fig:bias2} assumes that the
estimated frequency of the mode is the weighted combination of the
individual frequencies, where the weights are proportional to the
relative heights of the components, ${\cal E}(i)_{lm}$ (see
Equation~\ref{eq:vis}). We weight in this way because the estimated
frequency is influenced by the relative heights of the different
$m$. We then have:
 \begin{equation}
 \nu_{n2} =  \left( \sum_{m=-2}^{m=+2} {\cal E}(i)_{2m} \times \nu_{n2m} \right) / 
              \left( \sum_{m=-2}^{m=+2} {\cal E}(i)_{2m} \right).
 \label{eq:ass1}
 \end{equation}
The solid lines in Fig.~\ref{fig:bias2} show $\nu_{nl} - \nu_{\rm
cen}$, i.e., the predicted frequency bias, as a function of $i$. To
make the prediction we adopt $a_2=0.1\,\rm \mu Hz$ (typical value at
solar maximum). (The results do not depend upon $a_1$, which cancels
when $\nu_{nl}$ is computed.)

Results on Sun-as-a-star data, and simulated asteroseismic data on
solar-type stars, suggest that the weights are actually non-linear
functions in ${\cal E}(i)_{lm}$ (e.g., see Chaplin et al. 2004b). A
better description is one for which the weights are proportional to
${\cal E}(i)_{lm}$ raised to some positive power (four is probably a
reasonable value). Here, we therefore also make predictions based upon
 \begin{equation}
 \nu_{n2} =  \left( \sum_{m=-2}^{m=+2} {\cal E}(i)^4_{2m} \times \nu_{n2m} \right) / 
              \left( \sum_{m=-2}^{m=+2} {\cal E}(i)^4_{2m} \right),
 \label{eq:ass2}
 \end{equation}
which gives the dashed lines in Fig.~\ref{fig:bias2}.

Equations~\ref{eq:ass1} and~\ref{eq:ass2} may be written in more
general form (i.e., for any given $l$) as:
 \begin{equation}
 \nu_{nl} =  \left( \sum_{m=-l}^{m=+l} {\cal E}(i)^\gamma_{lm} \times \nu_{nlm} \right) / 
              \left( \sum_{m=-l}^{m=+l} {\cal E}(i)^\gamma_{lm} \right),
 \label{eq:ass3}
 \end{equation}
where $\gamma$ is a constant.


The results indicate that the bias is least severe at intermediate
values of $i$. The worst case is when $i=0$\,degrees. The non-linear
weighting in ${\cal E}(i)_{lm}$ also changes slightly the functional
form of the bias in $i$, e.g., in the case of $l=1$ it extends the
range of $i$ over which the bias is more important.

When we obtain long datasets on solar-type stars, frequency
uncertainties will drop to levels well below $0.1\,\rm \mu Hz$.  In
this context, the bias at some angles is by no means
insignificant. Moreover, the bias would of course be larger on a star
with higher-than-solar acoustic asphericity, i.e., one might quite
reasonably expect to encounter solar-type stars where the bias may be
a significant fraction of $1\,\rm \mu Hz$ or more. In sum, this source
of bias in the frequencies will need to be taken into account in the
not-too-distant future, if we are to avoid errors on the inferences
drawn on target stars.

 \subsection{Impact on frequency separation ratios}
 \label{sec:seprat}

Roxburgh \& Vorontsov (2003) proposed the use for asteroseismology of
the ratio of separations in frequency between low-$l$ p-modes:
\emph{Frequency separation ratios} are formed from the large and small
frequency separations of the p modes,
 \begin{equation}
 \Delta\nu_l(n) = \nu_{nl} - \nu_{n-1,l},
 \label{eq:dnu}
 \end{equation}
and
 \begin{equation}
 d_{l\,l+2}(n) = \nu_{n,l} - \nu_{n-1,l+2}.
 \label{eq:d}
 \end{equation}
Data on modes in the range $0 \le l \le 3$ may be used to make ratios,
according to:
 \begin{equation}
 r_{02}(n) = \frac{d_{02}(n)}{\Delta\nu_1(n)}, ~~~~~~~~r_{13}(n) =
 \frac{d_{13}(n)}{\Delta\nu_0(n+1)}.
 \label{eq:rats}
 \end{equation}
Roxburgh \& Vorontsov used an asymptotic method to show that the
ratios are very insensitive to conditions in the near-surface layers
of a solar-type star. This is an attractive property since the outer
layers are uncertain and hard to model accurately. As such, the ratios
in principle offer a clean diagnostic of the internal properties of a
star (see also Ot\'i Floranes et al. 2005 and Roxburgh 2005).  They
are nevertheless potentially sensitive to the acoustic asphericity --
as was pointed out by Ot\'i Floranes et al. (2005), and discussed at
length by Chaplin et al. (2005) -- which is clearly undesirable if one
wishes to suppress any signatures from the near-surface layers.

\begin{figure*}

\centerline
 {\epsfxsize=11.0cm\epsfbox{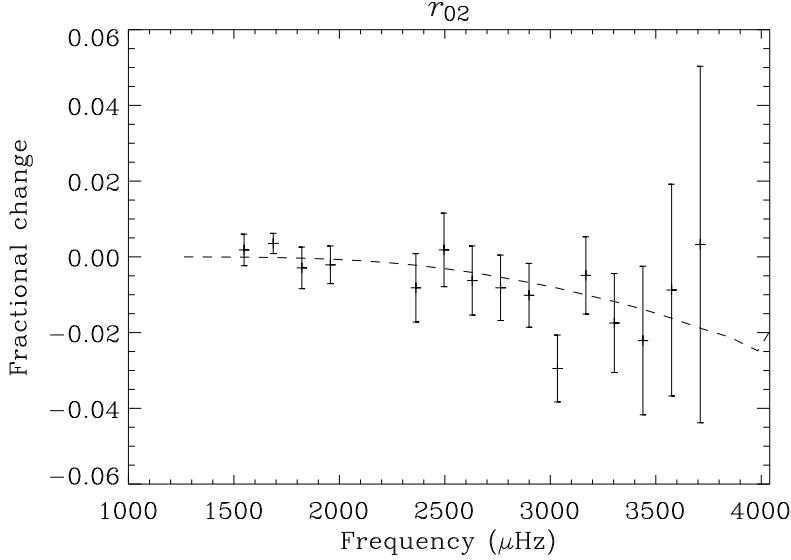}}

 \caption{Fractional change in $r_{02}$ -- as averaged over the most
   prominent low-$l$ modes -- between two BiSON Sun-as-a-star
   datasets, one around solar maximum and one around solar minimum
   (differences in sense maximum minus minimum). (From Chaplin et
   al. 2005.)}

 \label{fig:seprat1}

\end{figure*}

That the ratios are sensitive to the solar cycle may be understood as
follows. For the Sun-as-a-star data, measured frequency shifts of the
$l=2$ modes are larger than those of the neighbouring $l=0$ modes
(because the frequencies of these modes are dominated by the $|m|=2$
components, which are more sensitive to the acoustic perturbations
arising in the active latitudes than are the $l=0$ modes). The
measured $d_{l\,l+2}(n)$ will therefore carry a residual signature of
the solar cycle, being smaller at higher levels of activity. Residual
solar-cycle changes in $\Delta\nu_l(n)$ are, fractionally, much
smaller (but nevertheless measurable; see Broomhall et al. 2011), so
that in the case of, for example, $r_{02}(n)$:
 \begin{equation}
 \delta r_{02}(n) / r_{02}(n) \simeq \delta d_{02}(n)/
 d_{02}(n).
 \label{eq:dr}
 \end{equation}
For the Sun-as-a-star data, $\delta d_{02}(n)/ d_{02}(n) \approx
-0.01$ for modes at the centre of the p-mode spectrum observed at
solar maximum, hence, we would expect the ratios to decrease between
solar minimum and solar maximum by about this
amount. Fig.~\ref{fig:seprat1} plots measured fractional differences
in $r_{02}$, as averaged over the most prominent low-$l$ modes. The
differences were computed between two BiSON Sun-as-a-star datasets,
one formed of from around solar maximum, the other formed of data from
around solar minimum.


The potential impact on observations of other solar-type stars will
depend upon both the nature of the near-surface magnetic fields, and
also the angles of inclination offered by the stars. In
Section~\ref{sec:spat} below, we shall develop a simple model of the
frequency shifts that will allow us to make these predictions for
other stars. We therefore return in that section to the frequency
separation ratios.

 \subsection{Impact on shapes of resonant peaks}
 \label{sec:peak}

What effect will stellar-cycle shifts in frequency through a long
timeseries have on the underlying shapes of the p-mode peaks, when
those peaks are observed in frequency-power spectra made from the full
timeseries? Given a significant shift, one would expect the
Lorentzian-like peak shapes to be distorted, with the mode power being
spread out in frequency thereby flattening the profiles. Whether or not
a frequency shift $\delta\nu$ is significant in this context is
determined by the ratio 
 \begin{equation}
 \epsilon = \delta\nu/\Delta,
 \label{eq:epsi}
 \end{equation}
i.e., the ratio of the shift to the linewidth of the mode. The larger this
ratio, the more distorted a mode peak will be.

Chaplin et al. (2008b) derived analytical descriptions of the distorted
peak profiles. Since for solar-type stars we expect the modes to be
excited and damped on timescales much shorter than that on which any
significant change of the mode frequencies is observed, the profiles
may be assumed to correspond to the average of all the instantaneous
profiles taken at any time $t$ within the full period of observation
$T$. Each instantaneous profile may be described as, for example, a
Lorentzian with a central frequency $\nu(t)$$t$, such that the
time-averaged profile is
 \begin{equation}
 \label{eq:meanprofile}
 \left< P(\nu) \right> = \frac{1}{T}
 \displaystyle\int_{0}^{T} \frac{H}{1 + \left( \frac{\nu - \nu(t)}{\Delta/2}
 \right)^2}~dt,
 \end{equation}
where the angled brackets indicate an average over time, and $H$ and
$\Delta$ are the mode height (maximum power spectral density) and
linewidth, respectively.

Chaplin et al. (2008b) derived profiles given by two functions
describing the frequency shifts in time: first, the simplest possible
function, this being a linear variation over time (which would be
relevant for describing the profiles given by observations made on the
steepest parts of the rising or falling phases of a stellar cycle);
and second, a co-sinusoidal variation, to mimic a full stellar
cycle. They neglected the effects of the solar-cycle variations in
mode power, height and width.

\begin{figure}
 \centerline{\epsfxsize=10cm\epsfbox{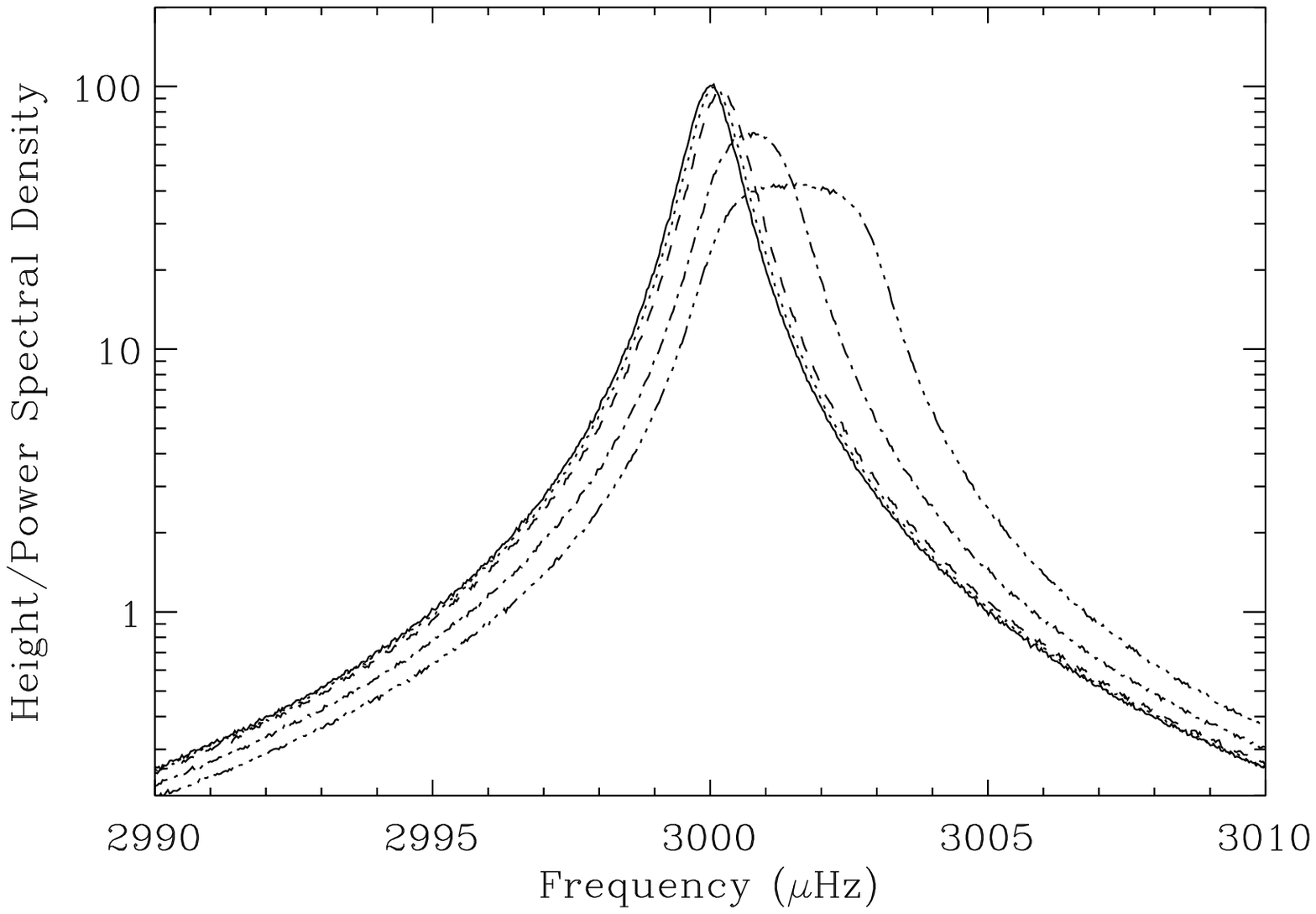}}
 \centerline{\epsfxsize=10cm\epsfbox{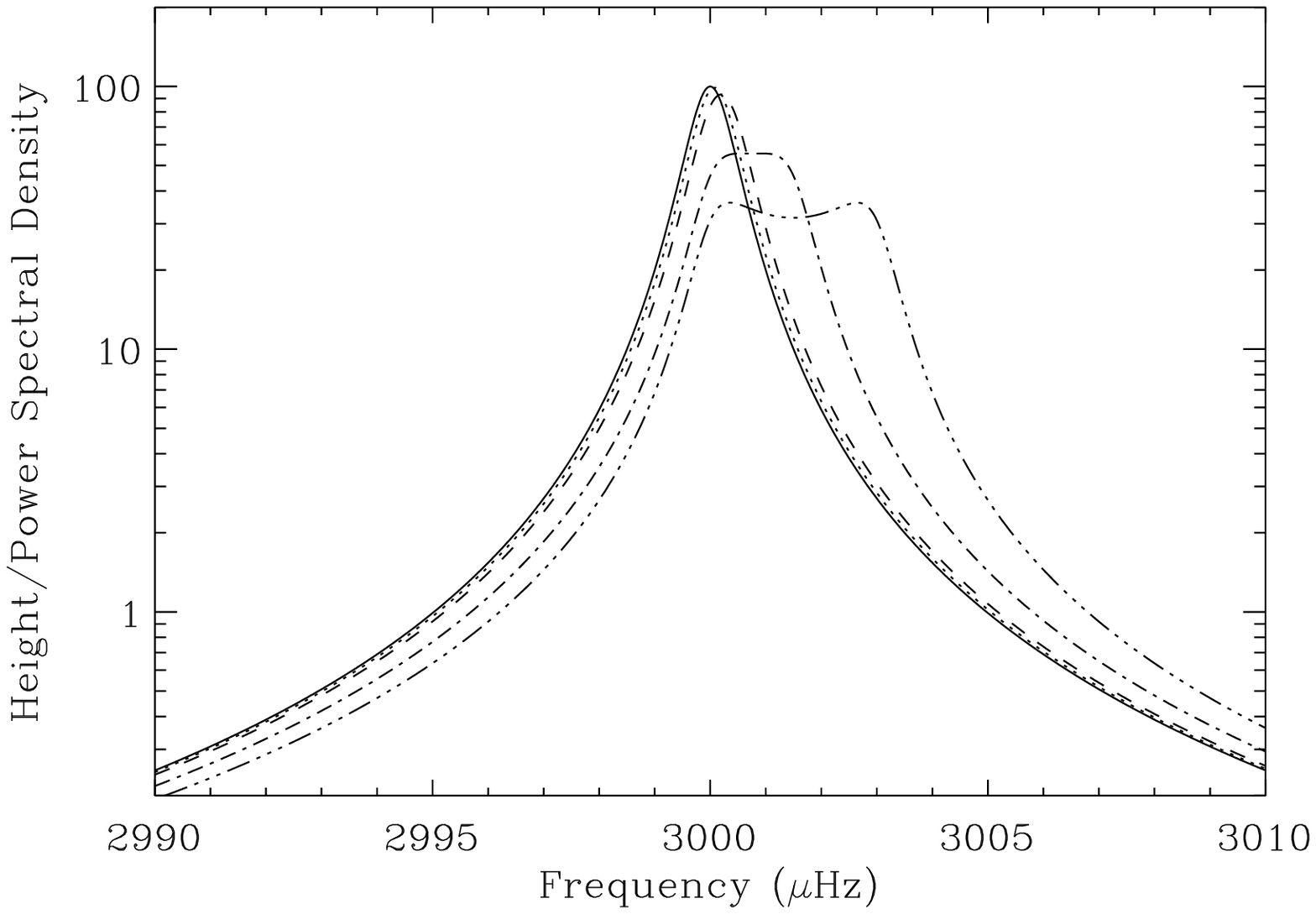}}

 \caption{Top panel: Peak profiles expected for a single mode of width
 $\Delta = 1\,\rm \mu Hz$ in the frequency-power spectrum of a time
 series within which the frequency was varied in a linear manner by
 total amount $\delta\nu$.  Various linestyles are: no shift (solid
 line); $\delta\nu=0.15\,\rm \mu Hz$ (dotted line); $0.40\,\rm \mu Hz$
 (dashed line); $1.50\,\rm \mu Hz$ (dot-dashed line); and $3.0\,\rm
 \mu Hz$ (dot-dot-dot-dashed line). Bottom panel: Peak profiles
 expected for a single mode of the same width, but where the frequency
 has instead been subjected to a co-sinusoidal variation in time.  The
 timeseries is assumed to have length equal to one cycle period; while
 the amplitude of the cycle is $\delta\nu/2$ (giving a total
 minimum-to-maximum shift in frequency of $\delta\nu$). Linestyles are
 as per the upper panel. (From Chaplin et al. 2008b.)}

 \label{fig:profs}
 \end{figure}

A simple linear variation in time may be described by
 \begin{equation}
 \label{eq:modefreq}
 \nu(t) = {\nu_0} + \delta \nu \frac{t}{T},
 \label{eq:lin}
 \end{equation}
where $\nu_0$ is the unperturbed frequency, and the frequency is
shifted by an amount $\delta\nu$ from the start ($t=0$) to the end
($t=T$) of the timeseries. Substitution of Equation~\ref{eq:lin}
into Equation~\ref{eq:meanprofile}, followed by solution of the
integral, gives the predicted mode profile:
 \begin{equation}
 \label{eq:linmeanprofile}
 \left<P(\nu) \right> = \frac{H}{2\epsilon} {\rm atan} \left( \frac{2 \epsilon}{1 - \epsilon^2 +
 X^2}\right),
 \end{equation}
where $\epsilon$ is the shift-to-linewidth ratio
(Equation~\ref{eq:epsi}), and
 \begin{equation}
 \label{eq:X}
 X= \frac{\nu - (\nu_0+ \delta\nu/2)}{\Delta/2}.
 \end{equation}
The top panel of Figure~\ref{fig:profs} shows profiles given by
Equation~\ref{eq:linmeanprofile}. The unperturbed profile (solid line)
is for a mode having an unperturbed frequency of $\nu_0=3000\, \rm \mu
Hz$, an unperturbed linewidth of $\Delta = 1\, \rm \mu Hz$, and an
unperturbed height of $H=100$ units. The other curves show the
profiles that result when the frequency shift, $\delta\nu$, is:
$0.15\,\rm \mu Hz$ (dotted line); $0.40\,\rm \mu Hz$ (dashed line);
$1.50\,\rm \mu Hz$ (dot-dashed line); and $3.0\,\rm \mu Hz$
(dot-dot-dot-dashed line). Since $\Delta = 1\, \rm \mu Hz$, the
$\delta \nu$ also correspond to the shift-to-linewidth ratios,
$\epsilon$. To put the values in context, we once more recall that
low-$l$ solar p modes at $\approx 3000\, \rm \mu Hz$, which also have
width $\approx 1\,\rm \mu Hz$, and show a frequency shift of
approximately $0.40\,\rm \mu Hz$ from the minimum to the maximum of
the solar activity cycle.

Only at the two largest shifts (dot-dashed and dot-dot-dot-dashed
lines) do the profiles depart appreciably from the Lorentzian
form. Closer inspection of the profiles does reveal some modest
distortion at the two, smaller, Sun-like shifts. These have $\epsilon
= 0.15$ and 0.40 respectively.


A co-sinusoidal variation of the mode frequency may be described by
 \begin{equation}
 \label{eq:solmodefreq}
 \nu(t) = {\nu_0} + \frac{\Delta\nu}{2}
 \left[ 1 - \cos\left(\frac{2\pi t}{P_{\rm
 cyc}}\right) \right],
 \end{equation}
where $\nu_0$ is again the unperturbed frequency (i.e., the frequency
at minimum activity), while $\delta \nu$ is now the full amplitude
(from minimum to maximum) of the cyclic frequency shift (not the
\emph{total} shift, as in the linear model).  When the length of
observation, $T$, equals the cycle period $P_{\rm cyc}$ (or one-half
of the period), the average profile resulting from
Equation~\ref{eq:solmodefreq} is:
 \begin{equation}
 \label{eq:cosmeanprofile}
 \left<P(\nu) \right> = \frac{\sqrt{L_1.L_2}} {\sqrt{1-\epsilon^2.F(L_1,L_2)}},
 \end{equation}
where
 \begin{eqnarray*}
 \label{eq:cosmeanprofileextra}
 F(L_1,L_2) & = & \frac{4 L_1 L_2}{H\left(\sqrt{L_1}+\sqrt{L_2}\right)^2},\\
 L_1 & = & \frac{H}{1+\left(X-\epsilon\right)^2},\\
 L_2 & = &\frac{H}{1+\left(X+\epsilon\right)^2},\\
 X &= & \frac{\nu - (\nu_0+\frac{\delta\nu}{2})} {\Delta/2}.
 \end{eqnarray*}
Since the frequency spends more time around its maximum and minimum
values, power near these extreme frequencies will have more weight in
the time-averaged profile, giving the profile a double-humped
appearance. This is reflected in the analytical expression through the
two Lorentzians $L_1$ and $L_2$.  It is obvious from
Equation~\ref{eq:cosmeanprofile} that when $\epsilon \ll 1$ the
profile tends to a single Lorentzian.

The bottom panel of Figure~\ref{fig:profs} shows predicted profiles
from Equation~\ref{eq:cosmeanprofile}\footnote{Versions of
Equations~\ref{eq:linmeanprofile} and~\ref{eq:cosmeanprofile}, which
allow for the small amounts of peak asymmetry observed in real solar p
modes, are presented in Chaplin et al. (2008). We note here that this
asymmetry is negligible for modes at the centre of the solar p-mode
spectrum.}, assuming observations made over a complete activity cycle,
and with the same shifts $\delta\nu$ as were applied in the
linear-model case (top panel of Figure~\ref{fig:profs}). As for the
simpler linear variation, it is only at the two largest $\delta\nu$
that the profiles depart appreciably from the unperturbed (Lorentzian)
form, here showing the predicted ``humps'' at the extreme frequencies
of the cycle. However, closer inspection again reveals some distortion
of the Lorentzian shapes at the small Sun-like shifts. For a given
shift, this distortion appears to be slightly larger than in the
simpler, linear case.

While the distortion seen at the Sun-like shifts is evidently modest,
Chaplin et al. (2008b) showed that it is nevertheless just detectable
in long, multi-year Sun-as-a-star datasets. By fitting modes to the
usual Lorentzian-like models -- which do not allow for the distortion
-- rather than modified models, like the ones shown above, Chaplin et
al. showed that an overestimation (underestimation) of the linewidth
(height) parameter results. This bias is estimated to be of size
comparable to the observational uncertainties given by datasets of
length several years. Bias in the frequency parameter is much less of
an issue. 

The distortion may of course be more important for asteroseismic
datasets on some solar-type stars, e.g., those for which the ratio of
the stellar-cycle frequency shifts to the mode linewidths is larger
than for the Sun. The shifts need only be about twice as strong as
those on the Sun before significant distortion of the peaks
results. Visible distortion of the mode profiles in asteroseismic data
may as such provide an initial diagnostic of strong stellar-cycle
signatures over the duration of the observations.

  \section{Estimated frequency shifts, and inference on spatial distribution of surface activity}
  \label{sec:spat}


 \begin{figure*}
\centerline {\epsfxsize=8.0cm\epsfbox{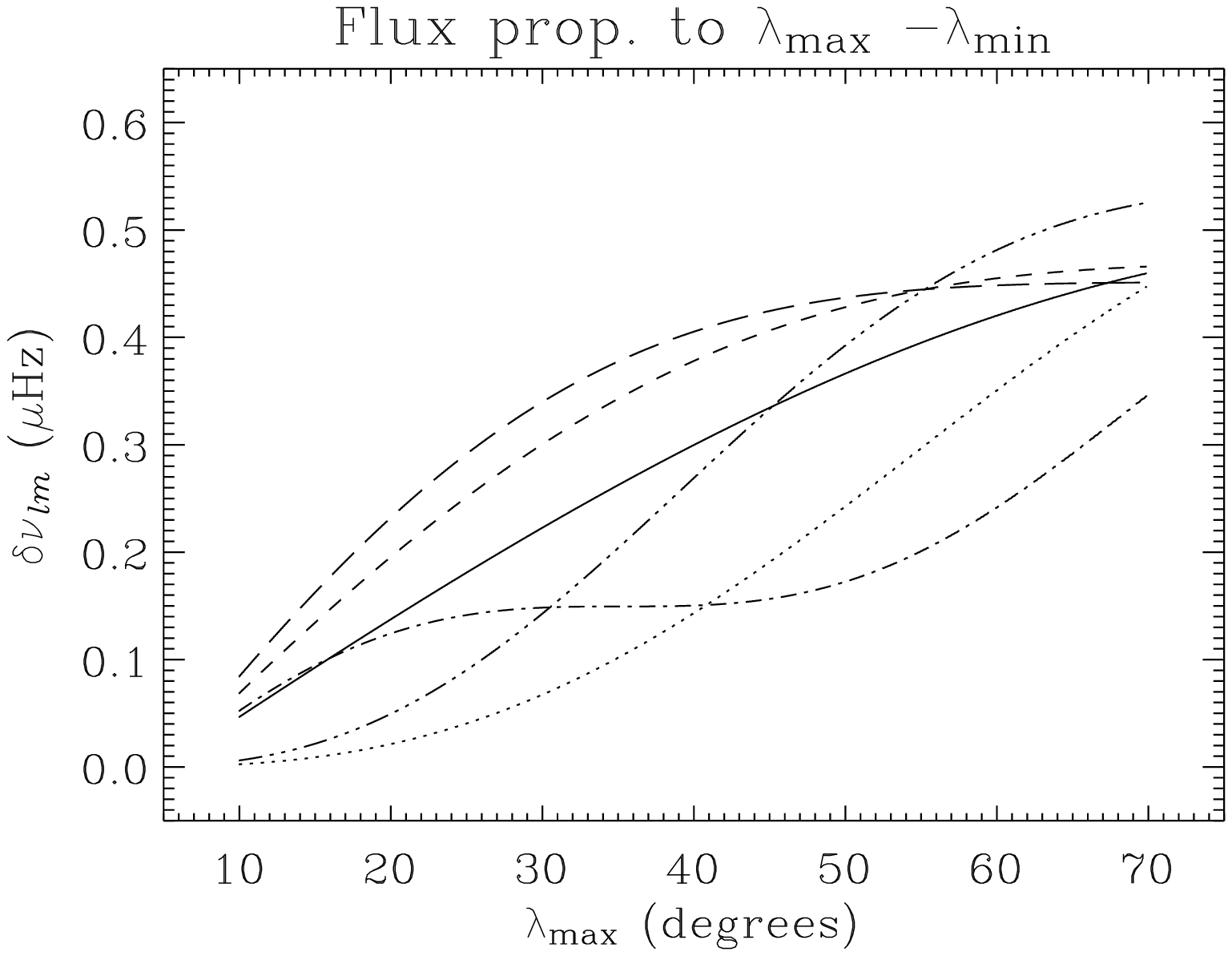}
 \epsfxsize=8.0cm\epsfbox{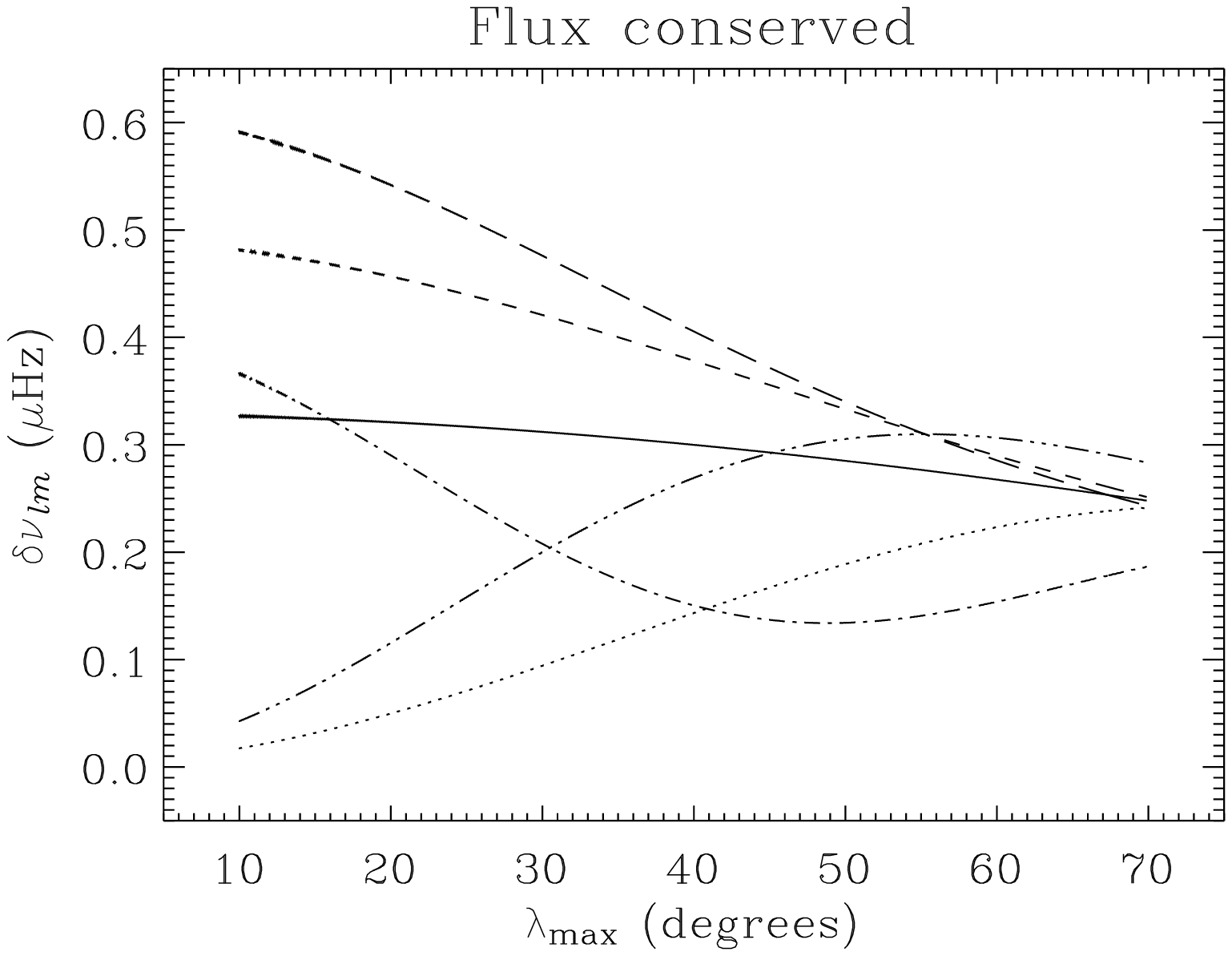}} \centerline
 {\epsfxsize=8.0cm\epsfbox{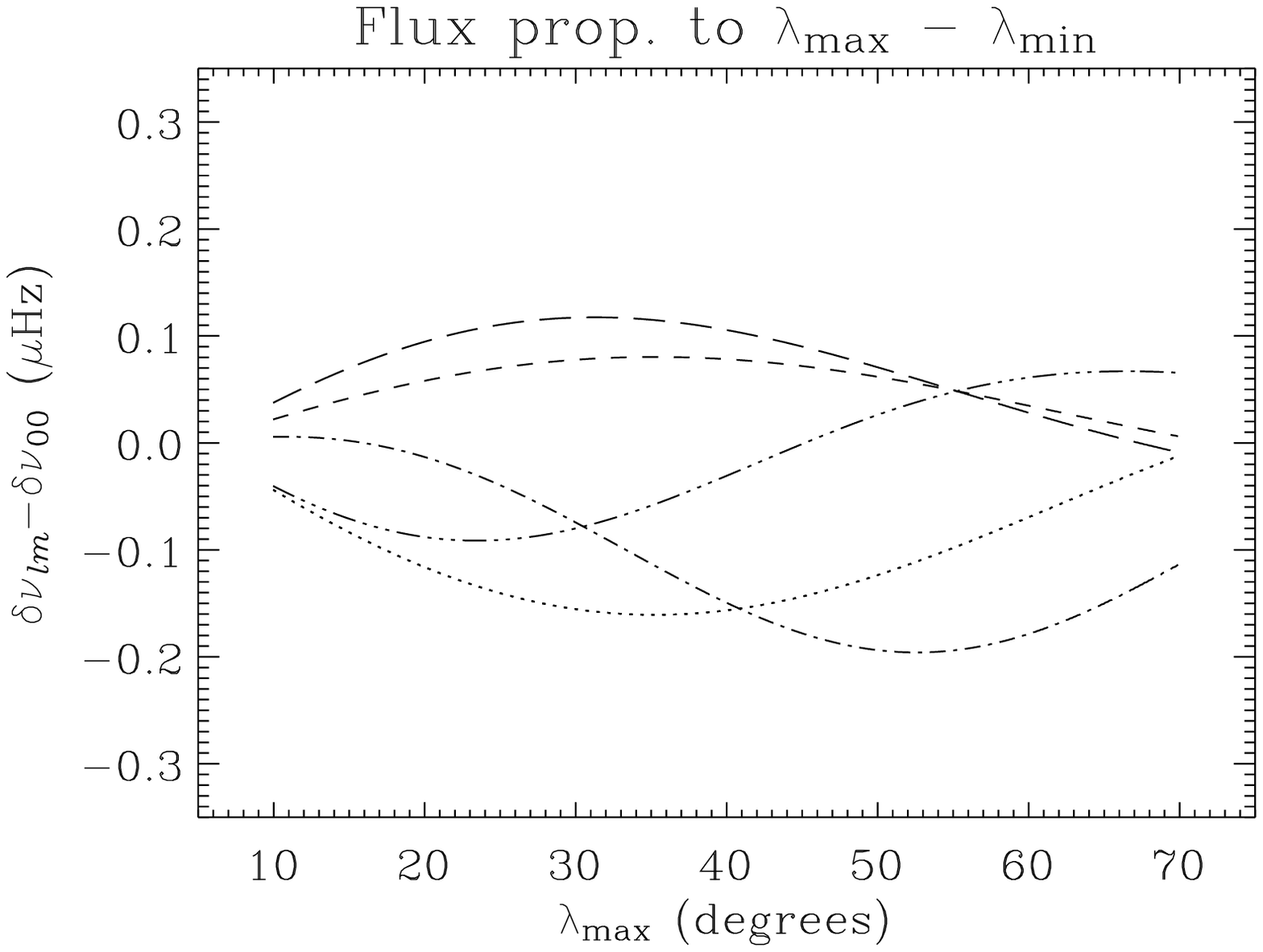}
 \epsfxsize=8.0cm\epsfbox{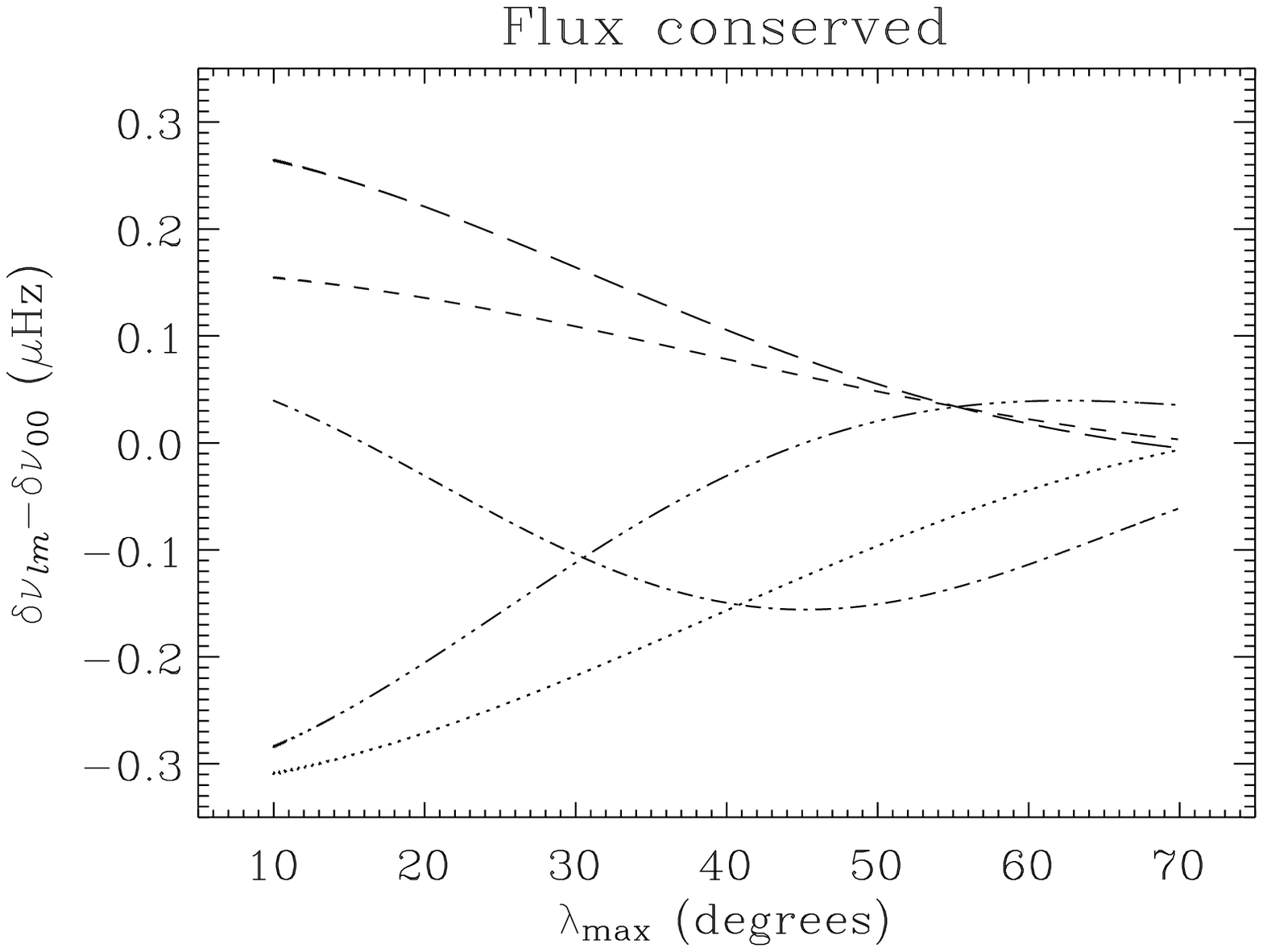}} \centerline
 {\epsfxsize=8.0cm\epsfbox{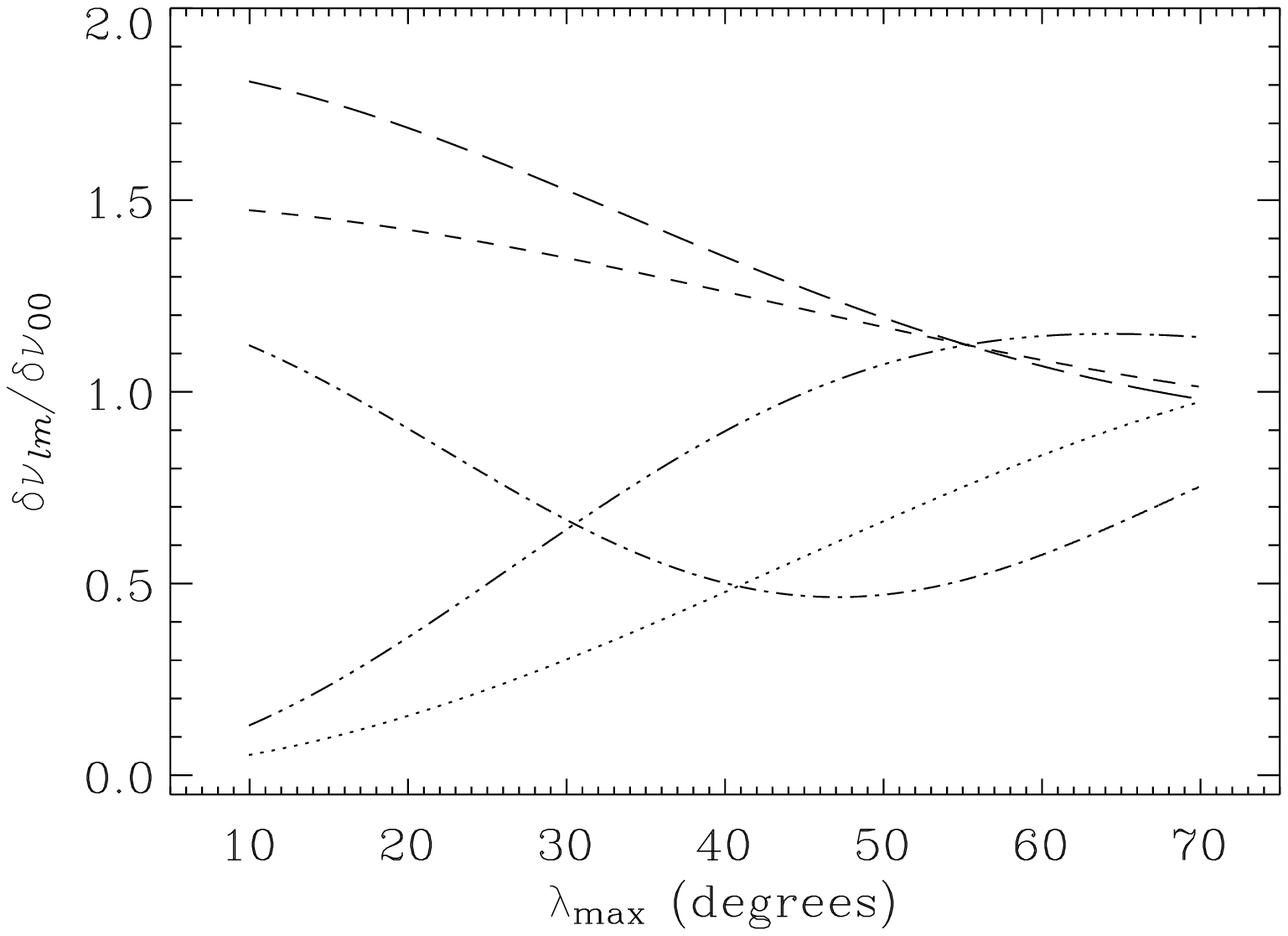}}

 \caption{Frequency shifts given by a simple model of stellar surface
 magnetic activity (magnetic flux). Linestyles, which are the same in
 all panels, show results for different ($l$,$|m|$) components as
 follows: solid for (0,0); dotted for (1,0); short dashes for (1,1);
 dot-dashed for (2,0); dot-dot-dot-dashed for (2,1); and long dashes
 for (2,2). Top panels: shifts calculated from models based on
 Equation~\ref{eq:mag1} (left-hand panel) and Equation~\ref{eq:mag2}
 (right-hand panel). Middle panels: residuals given by subtracting the
 (0,0) shifts from the other component shifts. Bottom panel: ratio of
 component shifts to (0,0) shifts (these ratios are the same for both
 sequences of models). (From Chaplin et al. 2007a.)}

 \label{fig:mis}
 \end{figure*}


We have seen that near-surface activity that is distributed in a
non-spherically symmetric manner (i.e., the aforementioned acoustic
asphericity) will give rise to differences in the magnitudes of the
frequency shifts of modes of different ($l,|m|$).  For a scenario like
that on the Sun, where bands of strong magnetic activity are located
at lower latitudes, it will be the sectoral components of the
non-radial modes that show the largest shifts. As was noted in
Section~\ref{sec:bias}, these components, for which $l=|m|$, have more
sensitivity to changes at lower latitudes than do their zonal
counterparts. It is possible to measure differences in the sizes of
the frequency shifts of the different low-$l$ modes (e.g., see Chaplin
et al. 2004a; Jim\'enez-Reyes 2004). If such measurements could be
made on other stars, might it not then be possible to make some
inference on the surface distribution in latitude of the activity,
from measurement of the relative sizes of the frequency shifts
(assuming, as mentioned elsewhere, that it is near-surface
perturbations that are the dominant cause of the observed shifts)?

This question was considered by Chaplin et al. (2007a).  They used a
very simple model of the spatial distribution of the surface activity
on a solar-type star, in which the activity was given a uniform
amplitude between some lower and upper bounds in latitude,
$\lambda$. As should be evident from what follows, it would be very
straightforward to incorporate a more sophisticated model of the
activity. They used this model to predict expected frequency shifts of
modes of different ($l,|m|$). Here, we reproduce their model, and use
it to show explicitly how it might be used with results on observed
frequency shifts to make inference on the active latitudes of a
star. Before that, we use its predictions to show how estimated
\emph{average} frequency shifts, and estimated frequency separation
ratios, of solar-type stars are affected by the angle of inclination.

Implicit in the calculation of the expected frequency shifts is the
assumption that contributions to the fractional change in the sound
speed are non-zero only very close to the surface. Consideration of
the radial dependence of the sound speed is neglected (again, to
simplify the description). The expected shifts, $\delta\nu_{lm}$ are
then proportional to (e.g., Moreno-Insertis \& Solanki 2000)
 \begin{equation}
 \delta \nu_{lm} \propto \left(l + \frac{1}{2} \right) \frac{(l-m)!}{(l+m)!}
 \int_{0}^{\pi} |P^m_l(\cos \theta)|^2 B(\theta) \sin \theta\,
 {\rm d}\theta,
 \label{eq:dnumis}
 \end{equation}
where the $P^m_l(\cos \theta)$ are Legendre polynomials, and
$B(\theta)$ describes the distribution in co-latitude $\theta$ of the
activity. (Note that $\lambda = 90\,{\rm degrees} - \theta$.) Chaplin
et al.  calibrated the model so that for a Sun-like configuration the
predicted $\delta\nu_{00}$ was equal to the observed average shift of
the solar $l=0$ modes. This calibration was subsumed within the
$B(\theta)$, as discussed below. For all surface configurations
tested, the activity at cycle minimum was set to $B(\theta)=0$.
Values of $\delta \nu_{lm}$ determined at the simulated cycle maxima
therefore corresponded to the sought-for stellar cycle shifts.

In a first sequence of models $B(\theta)$ was set to some constant
value at cycle maximum. This value was the same for \emph{all} model
configurations. The only parameter that was changed from model to
model, and therefore the only factor that could alter the observed
shifts, was the maximum latitude of the activity, $\lambda_{\rm
max}$. The minimum latitude was fixed for all models at $\lambda_{\rm
min} = 5\,\rm degrees$. In summary:
\begin{equation}
  \mbox{$B(\theta)=$} \left\{ \begin{array}{lll}
                 \mbox{const}&
                 \mbox{for $\lambda_{\rm min} \le |\lambda| \le \lambda_{\rm max}$},\\
                 \\ \mbox{$0$}&
                 \mbox{otherwise}.
                      \end{array}
                 \right.
\label{eq:mag1}
\end{equation}
In the scenario above, the total surface magnetic flux is proportional
to $\lambda_{\rm max} - \lambda_{\rm min}$, and therefore varied by
over an order of magnitude for the range of computations made ($10 \le
\lambda_{\rm max} \le 70\,\rm degrees$). Chaplin et al. also made a
second sequence of models in which the total flux was conserved for
all values of $\lambda_{\rm max}$. This model is described by:
\begin{equation}
  \mbox{$B(\theta)=$} \left\{ \begin{array}{lll}
                 \mbox{${\rm const}/ \left( \lambda_{\rm max}-\lambda_{\rm min} \right)$}&
                 \mbox{for $\lambda_{\rm min} \le |\lambda| \le \lambda_{\rm max}$},\\
                 \\ \mbox{$0$}&
                 \mbox{otherwise}.
                      \end{array}
                 \right.
\label{eq:mag2}
\end{equation}
Chaplin et al. used results from Knaack et al. (2001) to help guide
their calibration of Equations~\ref{eq:mag1} and~\ref{eq:mag2}. Knaack
et al. noted that observations of the surface activity during the
cycle~22 maximum showed sunspots confined to bands from 5 to
30\,degrees, and faculae in bands from 5 to 40\,degrees. On the Sun,
faculae occupy a much larger surface area than the spots (see also
De~Toma et al. 2004). The constant in Equations~\ref{eq:mag1}
and~\ref{eq:mag2} was therefore calibrated so that when $\lambda_{\rm
max} = 40\,\rm degrees$, the calculated shift of the radial modes was
$\delta \nu_{00} \sim 0.3\,\rm \mu Hz$. This is the value observed for
the most prominent low-$l$ radial modes of the Sun.

The top two panels of Fig.~\ref{fig:mis} show the calibrated shifts
from both sequences of models, based on Equation~\ref{eq:mag1}
(left-hand panel) and Equation~\ref{eq:mag2} (right-hand panel). The
middle panels show residuals given by subtracting the
($l,|m|$)$=$(0,0) shifts from the other component shifts. The bottom
panel shows the ratio of the component shifts and the (0,0) shifts:
since $B(\theta)=\rm const.$ for both sequences of models, the ratios
are also the same (hence only one plot).

Let us first consider the impact of the angle of inclination, $i$, on
estimated \emph{average} frequency shifts for solar-type stars. We
again adopt the approach of Section~\ref{sec:bias}, to take into
account the effect of the relative visibility of components within any
given $l$ (Equations~\ref{eq:ass1} to~\ref{eq:ass3}). We take the
frequency shifts $\delta\nu_{lm}$ predicted by the simple model in
this section, for different angles of inclination, $i$ (as computed
assuming the solar calibration). Since the $\delta\nu_{lm}$ are
functions of $i$, we write $\delta\nu_{lm}(i)$. We weight the shifts
of each component within a multiplet, to give the effective (weighted
average) shift at that $l$:
 \begin{equation}
 \delta\nu_{l} =  \left( \sum_{m=-l}^{m=+l} {\cal E}(i)^\gamma_{lm} \times
              \delta \nu_{lm} \right) / 
              \left( \sum_{m=-l}^{m=+l} {\cal E}(i)^\gamma_{lm} \right),
 \label{eq:nuall1}
 \end{equation}
Since one may then choose to average results over different $l$, to
reduce errors, we compute a final, weighted average shift according
to:
 \begin{equation}
 \left< \delta\nu \right> =  \left( \sum_{l=0}^{l=2} {\cal E}'_{l} \times
              \delta \nu_{lm} \right) / 
              \left( \sum_{l=0}^{l=2} {\cal E}'_{l} \right),
 \label{eq:nuall2}
 \end{equation}
using the predictions for $l=0$, 1 and 2. Here, the ${\cal E}'_{l}$
are the relative total $l$ visibilities, which we take to be 1, 1.5
and 0.5 (for $l=0$, 1 and 2, respectively).

\begin{figure*}

\centerline
 {\epsfxsize=11.0cm\epsfbox{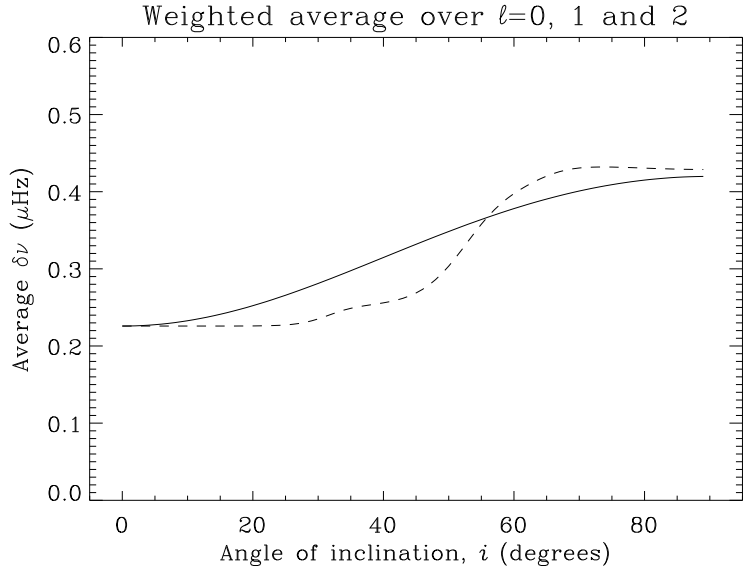}}

 \caption{Predicted frequency shift, averaged over $l=0$, 1 and 2
 modes, as a function of angle of inclination, $i$, assuming a
 solar-like spatial distribution of surface activity (see text for
 details).}

 \label{fig:nuall1}

\end{figure*}

Fig.~\ref{fig:nuall1} plots the $\left< \delta\nu \right>$ as a
function of $i$, for $\gamma=1$ (solid line) and $\gamma=4$ (dashed
lines). When $i$ is low, the $m=0$ components have the highest
visibility in the $l=1$ and $l=2$ modes, which reduces the size of the
average frequency shift (since these components show smaller shifts
than the other components when $\lambda_{\rm max} = 40\,\rm
degrees$). The plot shows clearly that for a solar-like spatial
distribution of surface activity, estimated average frequency shifts
could differ by up to a factor of two, depending upon the angle
$i$. It will clearly be important to have good constraints on $i$ in
order to properly interpret measured frequency shifts of solar-type
stars (e.g., see Ballot et al. 2006, 2008).


\begin{figure*}

\centerline
 {\epsfxsize=11.0cm\epsfbox{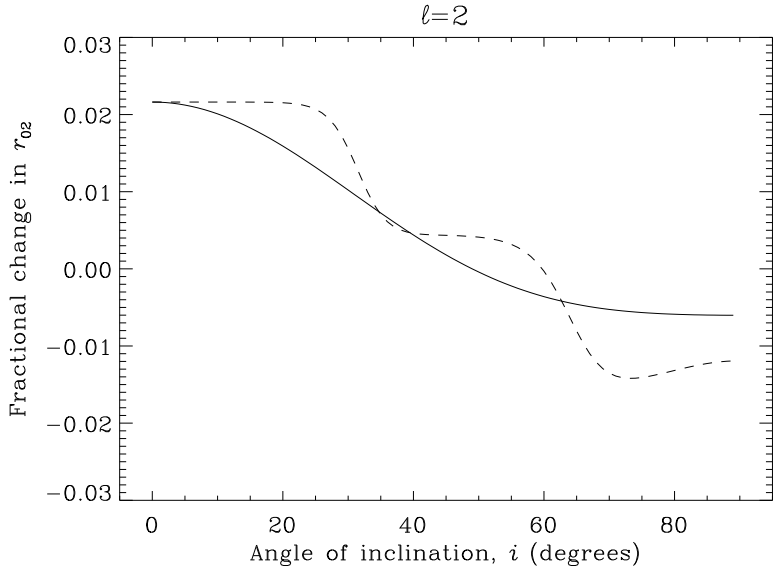}}

 \caption{Predicted fractional change (from solar minimum to solar
 maximum) in frequency separation ratios $r_{02}$ of the most
 prominent solar p modes, as a function of angle of inclination, $i$,
 assuming a solar-like spatial distribution of surface activity (see
 text for details).}

 \label{fig:seprat2}

\end{figure*}

We also use predictions from Equation~\ref{eq:nuall1} above to show
the impact of $i$ on the observed frequency separation
ratios. Fig.~\ref{fig:seprat2} plots the implied fractional change,
between solar minimum and maximum, in the frequency separation ratios
$r_{02}$ of the most prominent solar p modes. Recall from
Section~\ref{sec:seprat} that results from BiSON data give a
fractional change of $-0.01$; this is consistent with the prediction
shown here for $i=90\,\rm degrees$. Fig.~\ref{fig:seprat2} shows that
for the Sun the largest offset occurs at $i=0\,\rm degrees$, when the
predicted fractional change is 0.02 (positive). Since we should expect
to observe solar-type stars with larger acoustic asphericity than the
Sun, it is possible that bias in the frequency separation ratios could
reach levels of several per cent.


Let us now use the frequency shift \emph{ratios} plotted in the bottom
panel of Fig.~\ref{fig:mis}, together with results on measured
frequency shifts of low-$l$ solar p modes, to make an estimate of
$\lambda_{\rm max}$ for the Sun. By using the ratios as opposed to the
absolute values of the shifts we remove any dependence of the result
on the absolute calibration of the model shifts. Let us denote the
shift ratios by $k_{lm} = \delta\nu_{lm} / \delta\nu_{00}$. From the
BiSON Sun-as-a-star data in Chaplin et al. (2004a), we may infer
observed ratios of $k_{11} = 1.23 \pm 0.14$ and $k_{22} = 1.45 \pm
016$. (The observed $l=2$ Sun-as-a-star shifts will include a
contribution from the (2,0) components, however the contribution is
relatively weak, and the observed shifts may be regarded, to very good
approximation, as being those of the (2,2) components.) To infer
$\lambda_{\rm max}$, we use the appropriate curves from
Fig.~\ref{fig:mis}. Uncertainties in $\lambda_{\rm max}$ follow from
 \begin{equation}
 \delta\lambda_{\rm max} = \frac{\Delta \lambda_{\rm max}}{\Delta
 k_{\,lm}} \delta k_{\,lm},
 \label{eq:errr}
 \end{equation}
where $\delta k_{\,lm}$ are the observed uncertainties in the
$k_{lm}$, and $\Delta \lambda_{\rm max}/ \Delta k_{\,lm}$ are the
gradients of the respective curves in the vicinity of $k_{lm}$. From
the $l=1$ shift ratio we find $\lambda_{\rm max}=43 \pm 16\,\rm
degrees$, and from the $l=2$ shift ratio we find $\lambda_{\rm max}=34
\pm 10\,\rm degrees$. These values are of course reasonable estimates
for the Sun.

One may adopt a similar procedure using asteroseismic data on other
solar-type stars. One would again need to have good constraints on
$i$, which may be obtained from peak-bagging of the modes in the
frequency-power spectrum.

 \section{How long will it take to detect evolutionary changes?}
 \label{sec:evol}

We finish by thinking longer-term, and ask whether it
might be possible to measure evolutionary changes in solar p-mode
frequencies, using low-$l$ data (the choice of data again made with
other solar-type stars in mind).

\begin{figure*}
\centerline
 {\epsfxsize=11.0cm\epsfbox{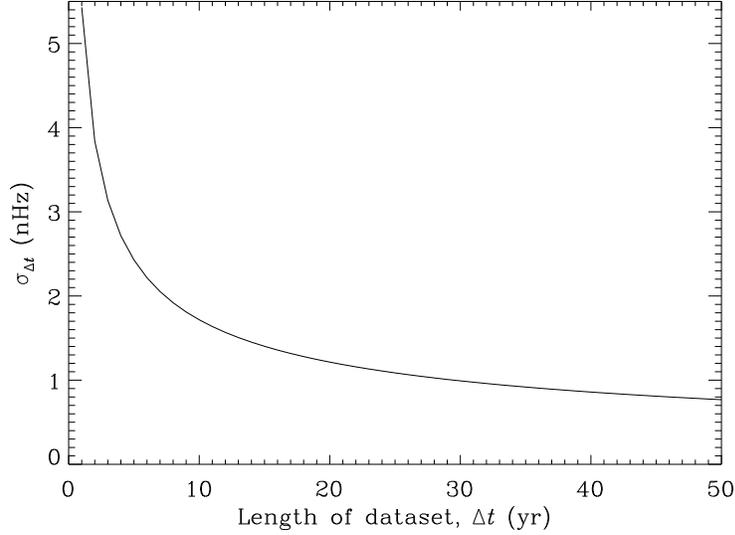}}

 \caption{Combined frequency precision, $\sigma_{\Delta t}$, for
 low-$l$ observations, extrapolated to different dataset lengths,
 $\Delta t$ (see text for more details).}

 \label{fig:evolve1}
 \end{figure*}

First, consider the precision in the estimated frequencies. We take
the estimated uncertainties, $\sigma_{nl}$, from Broomhall et
al. (2009), who measured low-$l$ frequencies in 8640\,d (23.7\,yr) of
BiSON Sun-as-a-star data. The published list included 81 frequencies,
covering $l=0$ to 3. An estimate of the \emph{combined} precision in
\emph{all} the measured frequencies, $\left< \sigma \right>$, is given
by:
 \begin{equation}
 \left< \sigma \right> = \left( \sum_{nl} 1/\sigma_{nl}^2
 \right)^{-1/2} \simeq 1.1\,\rm nHz.
 \label{eq:sigt1}
 \end{equation}
We may then estimate the combined precision, $\sigma_{\Delta t}$, for any dataset length,
$\Delta t$ (in yr) from: 
 \begin{equation}
 \sigma_{\Delta t} = \left( \frac{8640}{365} \right)^{1/2} \left< \sigma \right> \Delta
            t^{-1/2} \simeq 5.4\,\Delta t^{-1/2}\,\rm nHz,
 \label{eq:sigt2}
 \end{equation}
since the frequency uncertainties of the observed modes are expected
to scale with the square root of time. Fig.~\ref{fig:evolve1} plots
$\sigma_{\Delta t}$ as a function of dataset length, $\Delta t$.


\begin{figure*}
\centerline
 {\epsfxsize=11.0cm\epsfbox{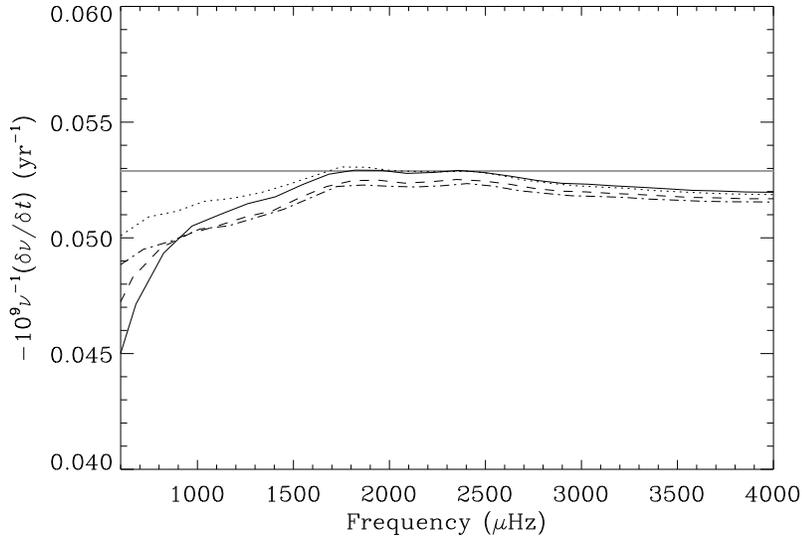}}

 \caption{$\/\nu (\delta\nu/\nu)$, as determined from the frequencies
  of several stellar evolutionary models made with solar mass and
  solar composition, but with varying ages, ranging $\pm 0.02\,\rm
  Gyr$ about the accepted solar age. Linestyles show data for $l=0$
  (thin solid line), $l=1$ (dotted line), $l=2$ (dashed line) and
  $l=3$ (dot-dashed line). The thick grey line plots $-(3/2R)(\delta
  R/R)$, as determined from the radii of the stellar models.}

 \label{fig:evolve2}
 \end{figure*}

How do the $\sigma_{\Delta t}$ compare to the expected frequency
shifts from evolutionary effects? Those shifts should, to good
approximation, be dominated by the slow expansion in radius of the Sun
as it evolves on the main sequence. Since, to first order, the
frequencies scale like the square root of the mean density of the
star, we have:
 \begin{equation}
 \left( \frac{\delta\nu}{\nu} \right) \simeq \frac{1}{2} \left(
 \frac{\delta \rho}{\rho} \right)
 \simeq -\frac{3}{2} \left( \frac{\delta R}{R} \right).
 \label{eq:evolve1}
 \end{equation}
This implies that
 \begin{equation}
 \frac{1}{\nu} \left( \frac{\delta\nu}{\delta t} \right) \simeq -\frac{3}{2R}
 \left( \frac{\delta R}{\delta t} \right).
 \label{eq:evolve2}
 \end{equation}
Fig.~\ref{fig:evolve2} plots $1/\nu (\delta\nu/\nu)$, as determined
from the frequencies of several stellar evolutionary models
(Yale-Yonsei models) made with solar mass and solar composition, but
with varying ages ranging $\pm 0.02\,\rm Gyr$ about the accepted solar
age. The different linestyles show results for different $l$ ($l=0$ as
a thin solid line, $l=1$ as a dotted line, $l=2$ as a dashed line, and
$l=3$ as a dot-dashed line).  The thick grey line plots
$-(3/2R)(\delta R/R)$, as determined from the radii of the stellar
models. It has a value $-5.3 \times 10^{-11}\,\rm yr^{-1}$, which
agrees quite well with the estimated gradient from the model-computed
frequencies, verifying that the frequencies scale in an approximately
homologous fashion.

We may then estimate to good approximation the expected evolutionary
frequency shift $\Delta \nu_{nl}$ of a mode over a length of time
$\Delta t$ (in yr):
 \begin{equation}
 \Delta\nu_{nl} \simeq - 5.3 \times 10^{-11} \nu_{nl} \Delta t.
 \label{eq:evolve3}
 \end{equation}
For a mode at the centre of the solar p-mode spectrum, having a
frequency $\approx 3000\,\rm \mu Hz$, we obtain $-1.6 \times
10^{-4}\,\rm nHz$ in 1\,yr, and $-1.6 \times 10^{-2}\,\rm nHz$ in
100\,yr. These expected shifts are significantly smaller than the
combined uncertainties expected for each length, being just over
5\,nHz in 1\,yr, and about 0.5\,nHz in 100\,yr. The shift and combined
uncertainty have similar sizes only when $\Delta t \approx 1100\,\rm
yr$.


\acknowledgements

The author would like to express his thanks to P.~Pall\'e and
colleagues at the IAC, for their invitation to lecture at the Winter
School and for the generous hospitality shown throughout the
meeting. He thanks the attendees and his fellow lecturers for making
the School such an enjoyable experience. He also acknowledges S.~Basu
for computing stellar model frequencies, A.-M. Broomhall and R.~Howe
for help with figures, and Y.~Elsworth for useful discussions.


\begin{thebibliography}{}

\bibitem[]{} Anguera Gubau, M., Pall\'e, P. L., Perez Hernandez,
F., R\'egulo, C. \& Roca-Cort\'es, T., 1992, A\&A, 255, 363

\bibitem[]{} Appourchaux, T., 1998, in: Structure and Dynamics
of the Interior of the Sun and Sun-like Stars SOHO 6/GONG 98 Workshop
Abstract, June 1-4, 1998, Boston, Massachusetts, p. 37

\bibitem[]{} Appourchaux, T., Michel, E., Auvergne, M., et al.\ 2008,
A\&A, 488, 705

\bibitem[]{} Appourchaux, T., Chaplin, W. J., 2007, A\&A, 469, 1151

\bibitem[]{} Ballot, J., Garc\'ia, R. A., Lambert, P., 2006, MNRAS,
  369, 1281

\bibitem[]{} Ballot, J., Appourchaux, T., Toutain, T., Guittet, M.,
  2008, A\&A, 486, 867

\bibitem[]{} Basri, G., Walkowicz, L. M., Batalha N., et al., 2010,
ApJ, 713, L155

\bibitem[]{} Basu, S., Mandel, A., 2004, ApJ, 617, L155

\bibitem[]{} B\"ohm-Vitense, E., 2007, ApJ, 657, 486

\bibitem[]{} Broomhall, A.-M., Chaplin, W. J., Elsworth, Y.,
  Fletcher, S. T., New, R., 2009, ApJ, 700, 162L

\bibitem[]{} Broomhall, A.-M., Chaplin, W. J., Elsworth, Y.,
  New, R., 2011, MNRAS, in the press

\bibitem[]{} Chaplin, W. J., Elsworth, Y., Isaak, G. R.,
Lines, R., McLeod, C. P., Miller, B. A., New, R., 1998, MNRAS, 300,
1077

\bibitem[]{} Chaplin, W. J., Appourchaux, T., Elsworth, Y.,
Isaak, G. R., Miller, B. A. \& New, R., 2000, MNRAS, 314, 75

\bibitem[]{} Chaplin, W. J., Appourchaux, T., Elsworth, Y.,
Isaak, G. R., New, R., 2001, MNRAS, 324, 910

\bibitem[]{} Chaplin, W. J.: 2004, in: SOHO 14/GONG 2004,
  `Helio- and Asteroseismology: Toward and Golden Future',
  ed. D. Dansey, ESA SP-559, Noordwijk, Netherlands, 34

\bibitem[]{} Chaplin, W. J., Elsworth, Y., Isaak, G. R.,
  Miller, B. A., New, R., 2004a, MNRAS, 352, 1102

\bibitem[]{} Chaplin~W.~J., Appourchaux~T., Elsworth~Y., Isaak~G.~R.,
Miller~B.~A., New~R., 2004c, A\&A, 424, 713

\bibitem[]{} Chaplin~W.~J., Appourchaux~T., Elsworth~Y., Isaak~G.~R.,
Miller~B.~A., New~R., Toutain, T., 2004b, A\&A, 416, 341

\bibitem[]{} Chaplin, W. J., Elsworth, Y., Miller, B. A.,
  New, R., Verner, G. A., 2005, ApJ, 635, 105

\bibitem[]{} Chaplin, W. J., Elsworth, Y., Houdek, G.,
  New, R., 2007a, MNRAS, 377, 17

\bibitem[]{} Chaplin, W. J., Elsworth, Y., Miller, B. A.,
  New, R., Verner, G. A., 2007b, ApJ, 659, 1749

\bibitem[]{} Chaplin, W. J., Houdek, G., Appourchaux, T., Elsworth,
  Y., New, R., Toutain, T., 2008a, A\&A, 483, 43

\bibitem[]{} Chaplin, W. J., Elsworth, Y., New, R., Toutain, T.,
  2008b, MNRAS, 384, 1668

\bibitem[]{} Chaplin, W. J., Appourchaux, T., Elsworth, Y., et al.,
  2010, ApJ, 713, L169

\bibitem[]{} Christensen-Dalsgaard~J., Berthomieu~G., 1991,
in: Solar Interior and Atmosphere, eds. A.~N.~Cox, W.~C.~Livingston,
M.~Matthews, Tucson, University of Arizona Press, p.~401

\bibitem[]{} De~Toma~G., White~O.~R., Chapman~G.~A.,
Walton~S.~R., Preminger~D.~G., Cookson~A.~M., 2004, ApJ, 609, 1140

\bibitem[]{} Elsworth, Y., Howe, R., Isaak, G. R., McLeod,
  New, R., 1990, Nat, 345, 322

\bibitem[]{} Elsworth, Y., Howe, R., Isaak, G. R., McLeod,
  Miller, B. A., New, R., Speake, C. C., Wheeler, S. J., 1994, ApJ,
  434, 801

\bibitem[]{} Dziembowski, W. A., Goode, P. R., 2005, ApJ, 625,
548

\bibitem[]{} Fletcher, S. T., Broomhall, A.-M., Salabert, D., Basu,
  S., Chaplin, W. j., Elsworth, Y., Garc\'ia, R. A., New, R., 2010,
  ApJ, 718, L19

\bibitem[]{} Garc\'ia, R. A., Mathur, S., Salabert, D., Ballot, J.,
  R\'egulo, C., Metcalfe, T. S., Baglin, A., 2010, Sci, 329, 1032

\bibitem[]{} Gelly B., Lazrek M., Grec G., Ayad A.,
Schmider F. X., Renaud C., Salabert D., Fossat E., 2002, A\&A, 394,
285

\bibitem[]{} Gilliland, R. L., Brown, T. M., Christensen-Dalsgaard,
  J., et al., 2010, PASP, 122, 131

\bibitem[]{} Gizon, L., Solanki, S. K., 2003, ApJ, 589, 1009

\bibitem[]{} Gonz\'alez-Hern\'andez, I., Howe, R., Komm, R., Hill, F.,
  2010, ApJ, 713, L16

\bibitem[]{} Gough, D. O., 1990, in: Progess of Seismology of the Sun
  and Stars, Proceedings of the Oji International Seminar Held at
  Hakone, Japan, eds. Y. Osaki, H. Shibahashi, Lecture Notes in
  Physics, vol. 367 (Springer-Verlag), p.~283

\bibitem[]{} Howe, R., Komm, R. W., Hill, F., 2002, ApJ, 580, 1172

\bibitem[]{} Howe, R., Christensen-Dalsgaard, J., Hill, F., Komm, R.,
  Schou, J., Thompson, M. J., 2009, ApJ, 701, L87

\bibitem[]{} Houdek, G., Chaplin, W. J., Appourchaux, T.,
  Christensen-Dalsgaard, J., D\"appen, W., Elsworth, Y., Gough, D. O.,
  Isaak, G. R., New, R., Rabello-Soares, M. C., 2001, MNRAS, 327, 483

\bibitem[]{} Jim\'enez-Reyes, S. J., R\'egulo, C., Pall\'e,
P. L., Roca-Cort\'es, T., 1998, A\&A, 329, 1119

\bibitem[]{} Jim\'enez-Reyes, S. J., Corbard, T., Pall\'e P. L.,
Roca-Cort\'es, T., Tomczyk, S., 2001, A\&A, 379, 622

\bibitem[]{} Jim\'enez-Reyes, S. J., Garc\'\i a, R. A.,
  Chaplin, W. J., Korzennik, S. G., 2004, ApJ, 610, 65

\bibitem[]{} Jim\'enez-Reyes, S. J., Chaplin, W. J., Elsworth,
Y., Garc\'\i a, R. A., Howe, R., Socas-Navarro, H., Toutain, T., 2007,
ApJ, 604, 1135

\bibitem[]{} Karoff, C., Metcalfe, T. S., Chaplin, W. J.,
  Elsworth, Y., Kjeldsen, H., Arentoft, T., Buzasi, D., 2009, MNRAS,
  399, 914

\bibitem[]{} Knaack~R., Fligge~M., Solanki~S.~K.,
Unruh~Y.~C., 2001, A\&A, 376, 1080

\bibitem[]{} Komm, R., Howe, R., Hill, F., 2000, ApJ, 531, 1094

\bibitem[]{} Libbrecht, K. G., Woodward, M. F., 1990, Nat, 345, 779

\bibitem[]{} Metcalfe, T. S., Dziembowski, W. A., Judge,
  P. G., Snow, M., 2007, ApJ, 723, 213

\bibitem[]{} Metcalfe, T. S., Basu, S., Henry, T. J.,
  Soderblom, D. R., Judge, P. G., Kn\"olker, M., Mathur, S., Rempel,
  M., 2010, ApJ, 723, 213

\bibitem[]{} Moreno-Insertis, F.~\& Solanki, S.~K.\ 2000, MNRAS, 313, 411

\bibitem[]{} Ot\'i Floranes~H., Christensen-Dalsgaard~J.,
Thompson~M.~J., 2005, MNRAS, 356, 671

\bibitem[]{} Roxburgh~I.~W., Vorontsov~S.~V., 2003, A\&A, 411,
215

\bibitem[]{} Roxburgh~I.~W., 2005, A\&A, 434, 665

\bibitem[]{} Pall\'e, P. L., R\'egulo, C., Roca-Cort\'es, T.,
1989, 224, 253

\bibitem[]{} Pall\'e, P. L., R\'egulo, C., Roca-Cort\'es, T.,
1990a, in: Progress of Seismology of the Sun and Stars, p. 129,
eds. Osaki Y. \& Shibahashi H., Springer-Verlag, Berlin

\bibitem[]{} Pall\'e, P. L., R\'egulo, C., Roca-Cort\'es, T.,
1990b, in: Progress of Seismology of the Sun and Stars, p. 189,
eds. Osaki Y. \& Shibahashi H., Springer-Verlag, Berlin

\bibitem[]{} Rabello-Soares, M. C., Korzennik, S. G., Schou, J., 2008,
  AdSpR, 41, 861

\bibitem[]{} Ritzwoller, M. H., Lavely, E. M., 1991, ApJ, 369, 557

\bibitem[]{} Salabert, D., Fossat, E., Gelly, B., Kholikov,
S., Grec, G., Lazrek, M., Schmider, F. X., 2004, A\&A, 413, 1135

\bibitem[]{} Salabert, D., Garc\'ia, R. A., Pall\'e, P.,
  Jim\'enez-Reyes, S. J., 2009, A\&A, 504, L1

\bibitem[]{} Sheeley, N. J., 2010, in: Proc. SOHO-23: Understanding a
  Peculiar Solar Minimum, eds. S. Crammer, T. Hoeksema \& J. Kohl,
  ASPCS, in the press (arXiv1005.3834v1)

\bibitem[]{} Tapping~K.~F., DeTracey~B., 1990, Sol Phys, 127, 321

\bibitem[]{} Tripathy, S. C., Jain, K., Hill, F., Leibacher, J. W.,
  2010, ApJ, 711, L84

\bibitem[]{} Toutain, T., Wehrli, C., 1997, in ASP
Conf. Ser. 118: 1st Advances in Solar Physics Euroconference. Advances
in Physics of Sunspots, ASP Conf. Ser. Vol. 118., Eds.: B. Schmieder,
J.C. del Toro Iniesta, M. Vazquez, p. 254.

\bibitem[]{} Verner, G. A., Chaplin, W. J., Elsworth, Y.,
  2006, MNRAS, 640, L95

\bibitem[]{} Watson, F., Fletcher, L., Dalla, S., Marshall, S., 2009,
  SolPhys, 260, 5

\bibitem[]{} Woodard, M. F., Noyes, R. W., 1985, Nat, 318,
449


\end{thebibliography}
\end{document}